\documentclass[aps,prd,onecolumn,nofootinbib,12pt]{revtex4-2}
\usepackage{amsmath,amssymb}

\usepackage{graphicx}  
\usepackage{dcolumn}   
\usepackage{bm}        
\usepackage[colorlinks=true,allcolors=blue]{hyperref}  
\usepackage{orcidlink}
\usepackage{slashed}
\usepackage{subcaption}
\usepackage{pgffor} 
\usepackage{subcaption}
\usepackage{booktabs}

\begin{document}

\title{Exploring twist-3 chiral even generalized parton distributions of light sea quarks in the proton using the light front model}

\author{Parashmani Thakuria\orcidlink{0000-0002-5652-7835}}
\email{parasht@tezu.ernet.in}
\affiliation{Department of Physics, School of Sciences, Tezpur University, Tezpur, India, Pin-784028}

\author{Madhurjya Lalung\orcidlink{0000-0002-7763-5050}}
\email{mlalung2016@gmail.com}
\affiliation{Department of Physics, Nagaon University, Nagaon, India}

\author{Jayanta Kumar Sarma}
\email{jks@tezu.ernet.in}
\affiliation{Department of Physics, School of Sciences, Tezpur University,\ Tezpur,\ India,\ Pin-784028\ }

\date{\today}

\begin{abstract}
We develop a light-front model of the proton to investigate the twist-3 chiral-even generalized parton distributions (GPDs) of light sea quarks. In this framework, sea quarks are treated as spin-\(\frac{1}{2}\) active partons, while the remaining proton constituents are modeled as spin-1 spectators. The light-front momentum wave function, derived from the soft-wall AdS/QCD approach, is fitted to unpolarized parton distribution functions (PDFs) from the CETQ global analysis to constrain the model parameters.
To further validate the model, we analyze the sea quark flavor asymmetry through the difference \(\bar{d}(x) - \bar{u}(x)\) and the ratio \(\bar{d}(x)/\bar{u}(x)\), achieving good agreement with known phenomenological trends. Using this fitted distribution, we compute the twist-3 chiral-even GPDs and examine their dependence on the transverse momentum transfer \(\boldsymbol{\Delta_T}\) and the longitudinal momentum fraction \(x\). In addition, we evaluate the twist-3 chiral-even PDF \(g_T(x)\) within this framework. The contribution of sea quarks to the proton's orbital angular momentum is also investigated and compared with existing results in the literature.

\textbf{Keywords}: Generalized parton distribution, QCD, Proton, Non-perturbative dynamics.
\end{abstract}

\maketitle

\section{Introduction}

It is an extremely challenging task to understand the internal three-dimensional structure of a proton in terms of its constituent partons (quarks, gluons)  \cite{deur2019spin,boffi2007generalized,kuhn2009spin,filippone2001spin,aidala2013spin,ji1997gauge}. With the help of parton distribution functions (PDFs), the one-dimensional function of the longitudinal momentum fraction and extensive study of the internal structures are carried out. PDFs are forward, diagonal matrix elements that provide a probabilistic interpretation for the operator chosen. Unlike PDFs, GPDs also depend on the squared momentum transfer ($t$) and the longitudinal momentum transfer (or skewness $\xi$). GPDs are off-forward matrix elements of non-local operators and can be accessed experimentally through processes like deeply virtual Compton scattering (DVCS) \cite{ji1997deeply,radyushkin1996scaling,belitsky2002theory,Hashamipour:2020kip} or deeply virtual meson production (DVMP) \cite{goloskokov2007longitudinal,goloskokov2008role,goloskokov2010attempt,goloskokov2011transversity}. Experimental data for these processes have been collected by collaborations like H1 \cite{adloff2002diffractive,aaron2009deeply}, ZEUS \cite{adloff2002diffractive,zeus2009measurement}, and fixed-target experiments at HERMES \cite{airapetian2012beam,airapetian2012beam1}, COMPASS \cite{d2004feasibility}, and JLab \cite{stepanyan2001observation,Goharipour:2024atx}. Considerable progress on the theoretical side has also been observed. Because of the perturbative nature of GPDs, they can't be directly calculated from the first principle of QCD. Although ongoing lattice calculations will provide more insights. Therefore, model calculations are carried out, such as the MIT bag model \cite{ji1997study}, the light-front constituent quark model \cite{boffi2007generalized,scopetta2004generalized,choi2002continuity,choi2001skewed}, the NJL model \cite{mineo2005generalized}, the
color glass condensate model \cite{goeke2008generalized}, the chiral quark-soliton model \cite{goeke2001hard,ossmann2005generalized}, the Bethe-Salpeter approach \cite{tiburzi2002exploring,noguera2004generalized} and
the meson cloud model \cite{pasquini2006virtual,pasquini2007generalized}.

Parton distribution functions (PDFs) are categorized by their twist, which refers to their order in a $1/Q$ expansion, where $Q$ denotes the hard scale associated with the underlying physical process. While most research on generalized parton distributions (GPDs) has focused on leading-twist contributions, higher-twist GPDs, particularly those of twist-3, remain comparatively under-explored. Nevertheless, twist-3 GPDs provide valuable insights into the partonic structure of hadrons. Notably, there is a connection between quark orbital angular momentum within the nucleon and twist-3 GPDs \cite{ji1997gauge,jaffe1990g1,hatta2012twist}. The amplitude of deeply virtual Compton scattering (DVCS) at twist-3 can be expressed in terms of twist-3 GPDs through the Compton form factor \cite{guo2022twist}. Some studies suggest that twist-3 GPDs can also offer insights into the average transverse color Lorentz force acting on quarks \cite{burkardt2013transverse},\cite{aslan2019transverse}. However, experimentally determining higher-twist distributions is challenging, as they are difficult to disentangle from leading-twist contributions. Moreover, higher-twist contributions are typically smaller, owing to suppression factors. The twist-3 GPDs have been investigated in the quark target model \cite{mukherjee2002off,mukherjee2003helicity,aslan2020singularities} and scalar diquark model \cite{aslan2020singularities} for the nucleon. The twist-4 proton GPDs were studied within the light-front quark-diquark
model (LFQDM) recently \cite{Sharma:2023ibp}, and the chiral-even twist-3 GPDs of the proton have also been investigated by the
Lattice QCD \cite{bhattacharya2023chiral}. The calculations of
the twist-3 GPDs chiral-even GPDs of the protons have been carried out in the basis light-front quantization (BLFQ collaboration) \cite{zhang2024twist} with the overlap representations of light-front wave functions.

In this work, we develop a light-front model of the proton in which the sea quark is treated as the active parton, while the remaining constituents are modeled as a spin-1 (axial-vector) spectator system. This modeling choice is motivated by both phenomenological observations and theoretical considerations. In the light-front Fock-state expansion of the proton, sea quark contributions emerge from higher Fock components such as \(|qqq\, q\bar{q}\rangle\), where a gluon splits into a quark–antiquark pair. By selecting the sea quark as the active parton from such a configuration, the residual system comprising the three valence quarks and the antiquark can effectively be approximated as a spin-1 spectator. This configuration naturally incorporates color and spin correlations and enables nontrivial helicity structures and orbital angular momentum couplings.

The inclusion of a spin-1 spectator is particularly suitable for modeling twist-3 generalized parton distributions (GPDs), especially in the chiral-even sector, where quark-gluon correlations and transverse spin dynamics are significant. The model also supports light-front wave functions derived from the soft-wall AdS/QCD approach, which encapsulates confinement effects and Regge behavior in an analytically tractable form. Notably, the spin-\(\frac{1}{2}\)–spin-1 structure enables mechanisms for generating sea quark flavor asymmetries, such as the empirically observed \(\bar{d}(x) > \bar{u}(x)\), through spin-dependent interference effects in the nonperturbative wave function.

The momentum wave functions employed in our analysis are motivated by the soft-wall AdS/QCD predictions. The model parameters are fixed by fitting the unpolarized sea quark parton distribution functions (PDFs) to the CT18NNLO global analysis at the scale \(Q_0 = 2\, \text{GeV}\). Using this framework, we investigate the behavior of twist-3 GPDs across a range of momentum fractions \(x\) and energy scales. In the forward limit, the only surviving chiral-even twist-3 PDF is \(g_T(x)\), which we examine in detail. We also analyze the validity of the Burkhardt--Cottingham sum rule~\cite{burkhardt1970sum}, providing insights into the integral properties of \(g_T\).

The paper is structured as follows. In Section~\ref{Sec:II}, we introduce the light-front model employed in this work, wherein the sea quark is treated as the active parton and the remaining constituents are modeled as a spin-1 spectator system. This section also includes a detailed discussion on the modeling of sea quark flavor asymmetries, such as \(\bar{d}(x) > \bar{u}(x)\), within our framework. In Section~\ref{Sec:III}, we present the overlap representation formalism for generalized parton distributions (GPDs), with a particular focus on the derivation of twist-3 chiral-even GPDs using light-front wave functions. Section~\ref{Results} is dedicated to presenting and analyzing the numerical results of our study. This includes a comparison of model predictions with experimental data and other theoretical approaches for  twist-3 GPDs and twist-3 parton distributions. The contribution of sea quarks to the proton's orbital angular momentum is also investigated and compared with existing results in the literature in this section. Finally, the conclusions are summarized in Section~\ref{Conclusion}, highlighting the implications of our findings and possible future directions.

\section{LIGHT-CONE MODEL DESCRIPTION}
\label{Sec:II}

We consider a two-particle light-front Fock state representation of the proton, denoted by \( |\lambda_q, \lambda_A, xP^+, \boldsymbol{k}_T\rangle \), where \( \lambda_q \) and \( \lambda_A \) represent the light-front helicities of the active sea quark and the spectator system, respectively. In our model, the active parton is a spin-\(\frac{1}{2}\) sea quark, and the spectator is treated as a spin-1 (axial-vector diquark) system. This spin configuration permits nontrivial helicity combinations and orbital angular momentum contributions, which are essential for modeling twist-3 GPDs.

Throughout our calculations, we adopt the light-front convention \( x^{\pm} = x^0 \pm x^3 \), and we work in a frame where the transverse momentum of the proton vanishes. In this frame, the four-momentum of the proton is given by
\begin{equation}
    P = \left(P^+, \frac{M^2}{P^+}, \boldsymbol{0}_T\right),
\end{equation}

where \( M \) denotes the proton mass.

The four-momentum of the active sea quark is parameterized as
\begin{equation}
    k_1 = \left(xP^+, k^-, \boldsymbol{k}_T\right),
\end{equation}
while that of the spectator is

\begin{equation}
k_2 = \left((1-x)P^+, k_X^-, -\boldsymbol{k}_T\right),
\end{equation}

ensuring overall momentum conservation in the transverse direction. Here, \( m \) denotes the mass of the constituent quark, and \( M \) is taken as \(0.938 \, \text{GeV}\). In nonperturbative QCD approaches such as Dyson–Schwinger equations, the constituent quark mass is consistently found to lie near
\(m\approx 0.30\text{–}0.35\,\mathrm{GeV}
\), emerging dynamically through chiral symmetry breaking in the proton wave function~\cite{Cloet2014}.

The two-particle Fock state for the proton with \( J^z = \pm \frac{1}{2} \) has six possible spin combinations for the quark and axial-vector diquark. It can be expressed as~\cite{Ellis_2009},
\begin{equation}
    |\bar{\nu} A \rangle^{\pm}= \sum_{\lambda_{\bar{\nu}}} \sum_{\lambda_A} \int \frac{dx \, d^2 \boldsymbol{k}_T}{2 (2\pi)^3 \sqrt{x(1-x)}} \, \psi^{\pm}_{\lambda_{\bar{\nu}}{\lambda_A}}(x,\boldsymbol{k}_T) \, |\lambda_{\bar{\nu}}, \lambda_A, xP^+, \boldsymbol{k}_T \rangle.
\end{equation}

 The two-particle state is represented by \( |\lambda_q, \lambda_A, xP^+, \boldsymbol{k}_T\rangle \), where the quark helicity is \( \lambda_{\bar{\nu}} = \pm \tfrac{1}{2} \), and the spectator axial-vector diquark helicity is \( \lambda_A = \pm1, 0 \) (triplet). In Table~\ref{table1}, we provide the light front wave functions (LFWFs) \cite{Ellis_2009,PhysRevD.95.074009} for the spin-1 spectator and spin-\(\frac{1}{2}\) active parton.

The momentum-space light-front wave function is motivated by predictions from the soft-wall AdS/QCD model for a system of a spin-1 spectator and a spin-\(\frac{1}{2}\) active parton~\cite{PhysRevD.95.074009}, and takes the form:
\begin{align}
\varphi_i^{\nu}(x,\boldsymbol{k}_T) = \frac{4\pi}{\kappa} \sqrt{\frac{\log\left(\frac{1}{1-x}\right)}{1-x}} \, x^{a_i^{\nu}} (1-x)^{b_i^{\nu}} \exp\left[-\delta_i^{\nu} \frac{\log\left(\frac{1}{1-x}\right)}{2\kappa^2 (1-x)^2} \, \boldsymbol{k}_T^2 \right],
\end{align}
where \(\nu\) represents the flavor of sea quark; \( \kappa = 0.4 \ \text{GeV} \) is the AdS/QCD scale parameter~\cite{chakrabarti2013generalized}, and \( a_i^{\bar{\nu}} \), \( b_i^{\bar{\nu}} \), and \( \delta_i^{\bar{\nu}} \) are model-dependent parameters. In our model, the value of \( \delta_i^{\bar{\nu}} \) is fixed at 1. The remaining parameters are determined by fitting the unpolarized sea quark parton distribution functions to the CT18NNLO global analysis at the scale \( Q_0 = 2 \ \text{GeV} \). These values are provided in table \ref{table2}.

\begin{table}[ht]
\caption{Wave function components $\psi_{\lambda_{\bar{q}} \lambda_A}^{J_z}$ for $J_z = \pm 1/2$.}
\label{table1}
\setlength{\tabcolsep}{10pt}
\begin{tabular}{llll}
\toprule
\textbf{$\lambda_{\bar{\nu}}$}&\textbf{$\lambda_{{A}}$}& \textbf{$J_z=+1/2$} & \textbf{$J_z=-1/2$} \\
\midrule
\rule{0pt}{4ex}%
+1/2 & +1 & $\psi_{++}^{+} = -\sqrt{\frac{2}{3}} \dfrac{-k_1+i k_2}{x} \varphi_2$ &
$\psi_{++}^{-} = 0$ \\
\rule{0pt}{4ex}%
-1/2 & +1 & $\psi_{-+}^{+} = +\sqrt{\frac{2}{3}} \dfrac{x M+m}{x} \varphi_1$ &
$\psi_{-+}^{-} = 0$ \\
\rule{0pt}{4ex}%
+1/2 & 0 & $\psi_{+0}^{+} = -\sqrt{\frac{1}{3}} \dfrac{x M +m}{x} \varphi_1$ &
$\psi_{+0}^{-} = -\sqrt{\frac{1}{3}} \dfrac{-k_1+i k_2}{x} \varphi_2$ \\
\rule{0pt}{4ex}%
-1/2 & 0 & $\psi_{-0}^{+} = +\sqrt{\frac{1}{3}} \dfrac{k_1+i k_2}{x} \varphi_2$ &
$\psi_{-0}^{-} = +\sqrt{\frac{1}{3}} \dfrac{x M+m}{x} \varphi_1$ \\
+1/2 & -1 & $\psi_{+-}^{+} = 0$ &
\rule{0pt}{4ex}%
$\psi_{+-}^{-} = -\sqrt{\frac{2}{3}} \dfrac{x M+m}{x} \varphi_1$ \\
-1/2 & -1 & $\psi_{--}^{+} = 0$ &
\rule{0pt}{4ex}%
$\psi_{--}^{-} = +\sqrt{\frac{2}{3}} \dfrac{k_1+i k_2}{x} \varphi_2$ \\
\bottomrule
\end{tabular}
\end{table}

The non-perturbative structure of hadrons is studied using parton distributions. Physically, it gives the probability of finding a parton with longitudinal momentum fraction $`x$'. The unpolarized sea quark PDF in our model is expressed as,

\begin{align}
    f_1(x)&=\int \frac{d^2 \boldsymbol{k}_T}{16 \pi^3} \bigg[|\psi_{++}^{+}(x,\boldsymbol{k}_T)|^2+|\psi_{-+}^{+}(x,\boldsymbol{k}_T)|+|\psi_{+0}^{+}(x,\boldsymbol{k}_T)|^2 +
     \notag \\ 
    &\hspace{4cm}|\psi_{-0}^{+}(x,\boldsymbol{k}_T)|^2+|\psi_{+-}^{+}(x,\boldsymbol{k}_T)|^2+|\psi_{--}^{+}(x,\boldsymbol{k}_T)|^2\bigg] \notag\\
    &=\int \frac{d^2 \boldsymbol{k}_T}{16 \pi^3} \bigg[|\psi_{++}^{-}(x,\boldsymbol{k}_T)|^2+|\psi_{-+}^{-}(x,\boldsymbol{k}_T)|+|\psi_{+0}^{-}(x,\boldsymbol{k}_T)|^2 
    + \notag \\ 
    &\hspace{4cm}|\psi_{-0}^{-}(x,\boldsymbol{k}_T)|^2+|\psi_{+-}^{-}(x,\boldsymbol{k}_T)|^2+|\psi_{--}^{-}(x,\boldsymbol{k}_T)|^2\bigg] \notag\\
    &=x^{2a_1-2}(1-x)^{2 b_1} \frac{\kappa^2(1-x)^2}{log\big[1/(1-x)\big]}+(x M+m)^2 x^{2 a_2-2}(1-x)^{2 b_2}.
\end{align}

\begin{figure}[htbp]
  \centering
  \begin{subfigure}[b]{0.47\textwidth}
    \includegraphics[width=\linewidth]{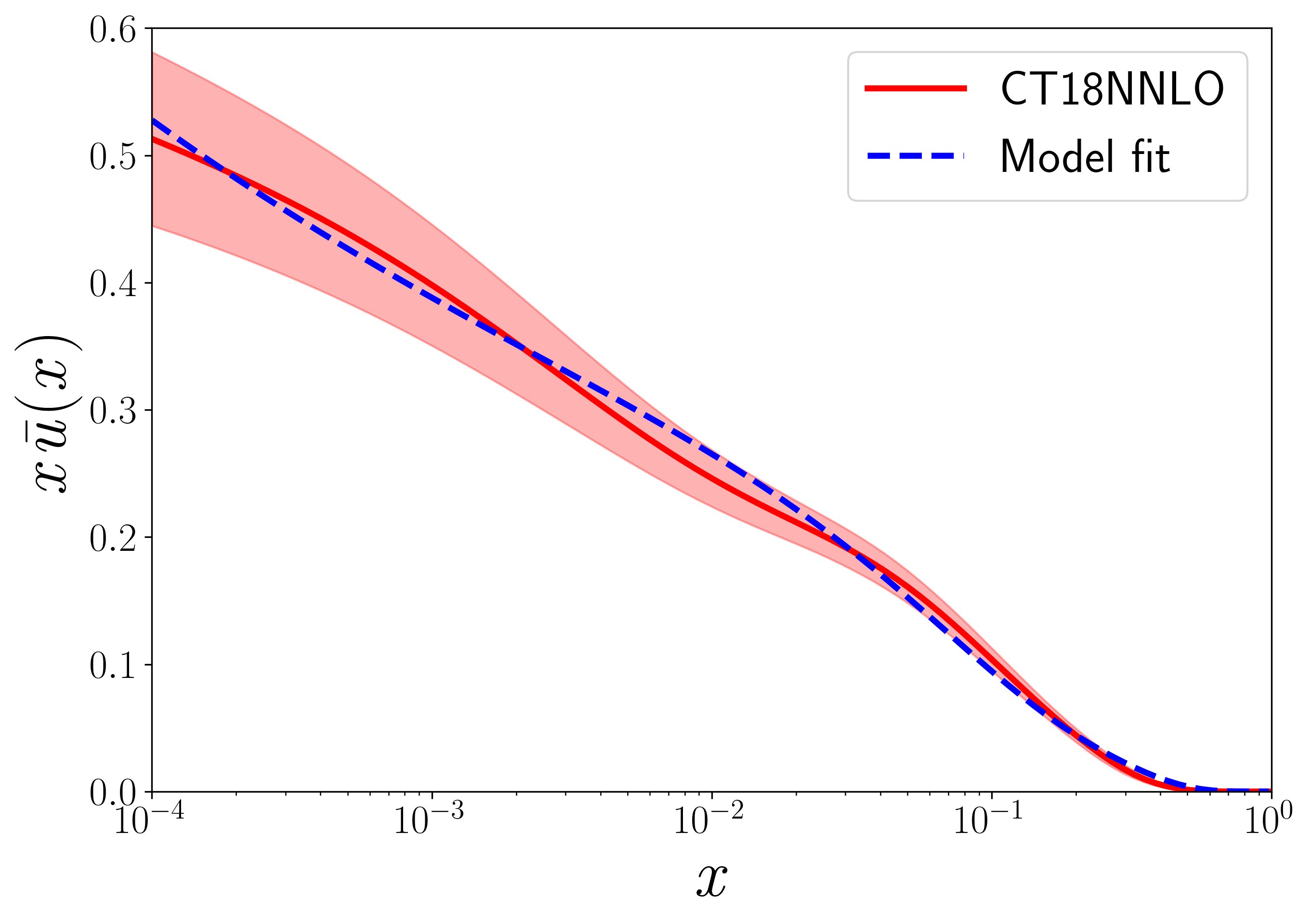}
    \caption{}
  \end{subfigure}
  \hspace{0.1cm}
  \begin{subfigure}[b]{0.47\textwidth}
    \includegraphics[width=\linewidth]{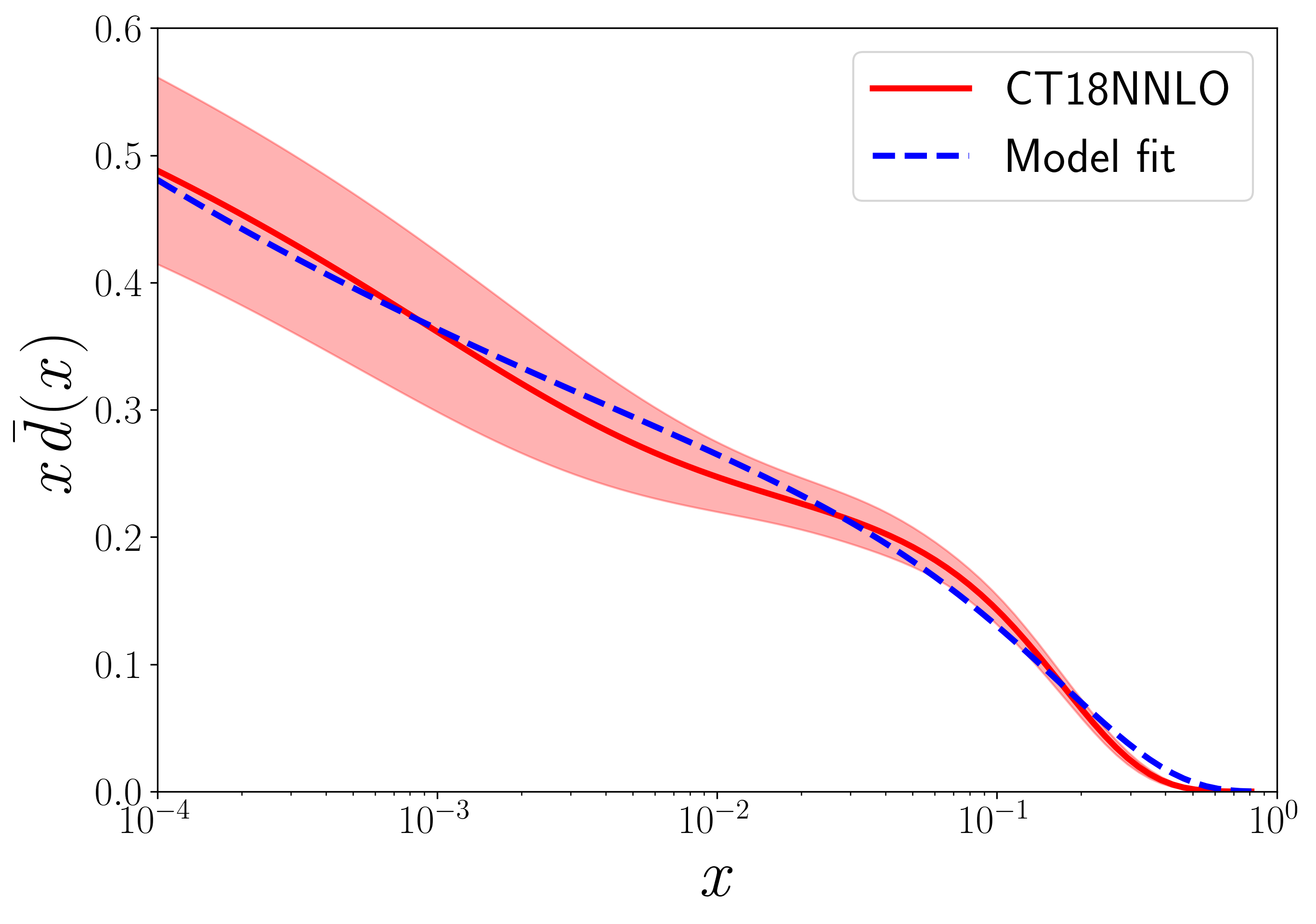}
    \caption{}
  \end{subfigure}
  \caption{Sea quark unpolarized PDFs in the proton are fitted to CT18NNLO data: panel \((a)\) shows the distribution for \(x \bar{u}\), and panel \((b)\) shows \(x \bar{d}\) in the kinematic region \(0.0001 < x < 1\) at the scale \(Q_0 = 2\, \text{GeV}\).}
  \label{fig1}
\end{figure}

We have followed the procedure provided by \cite{Choudhary:2023unw} to determine the model parameters. In our model, the parameters are $a_i^{\nu}$ and \(b_i^{\nu}\). The $a_i^{\nu}$ parameter controls the low $x$ value, and the $b_i^{\nu}$ parameter controls the high $x$ value. In fig \ref{fig1}, figure \((a)\) is for $x \bar{u}$ and figure \((b)\) is for $x \bar{d}$.

\begin{table}[hbt]
\caption{Fitted parameters for $\bar{u}$ and $\bar{d}$ quark distributions.}
\label{table2}
\setlength{\tabcolsep}{10pt}
\begin{tabular}{llllll}
\toprule
\textbf{} & \textbf{\(a_1\)}  & \textbf{\(b_1\)} & \textbf{\(a_2\)}  & \textbf{\(b_2\)} & \(\chi^2/\text{d.o.f}\) \\
\midrule

\textbf{\(\bar{u}\)} & \(0.9377 \pm0.0003\) & \(0.9769\pm0.034\) &\(2.5481\pm 0.0612\) &\(4.5576\pm 0.1028\) & 0.91\\

\textbf{\(\bar{d}\)} & \(1.1064\pm 0.0031\) & \(0.4092\pm0.0006\) &\(2.5723\pm0.0906\) &\(4.4223\pm0.0474\) & 0.67  \\
\bottomrule
\end{tabular}
\end{table}

\begin{figure}[htbp]
  \centering
  \begin{subfigure}[b]{0.47\textwidth}
    \includegraphics[width=\linewidth]{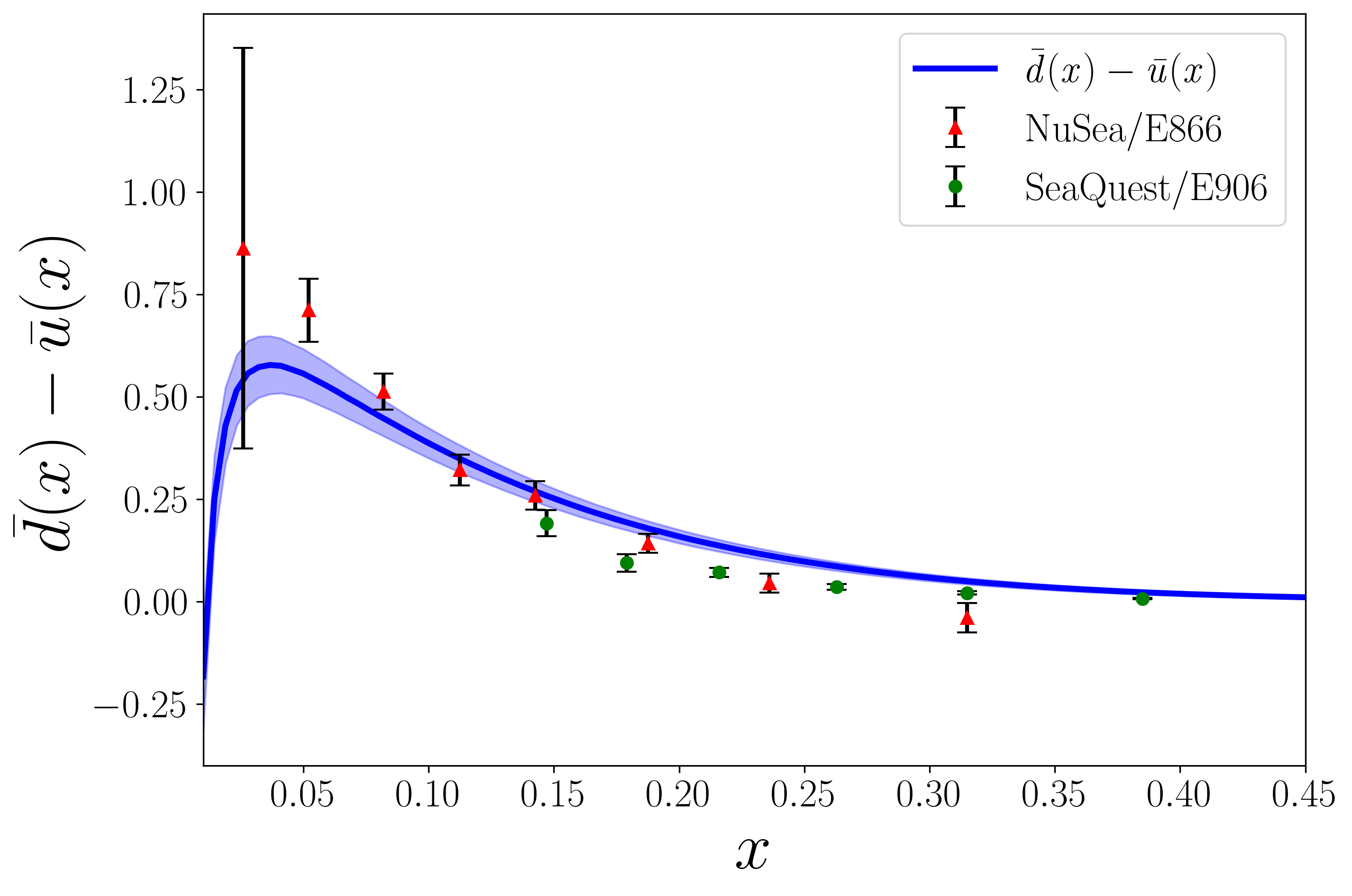}
    \caption{}
  \end{subfigure}
  \hspace{0.1cm}
  \begin{subfigure}[b]{0.46\textwidth}
    \includegraphics[width=\linewidth]{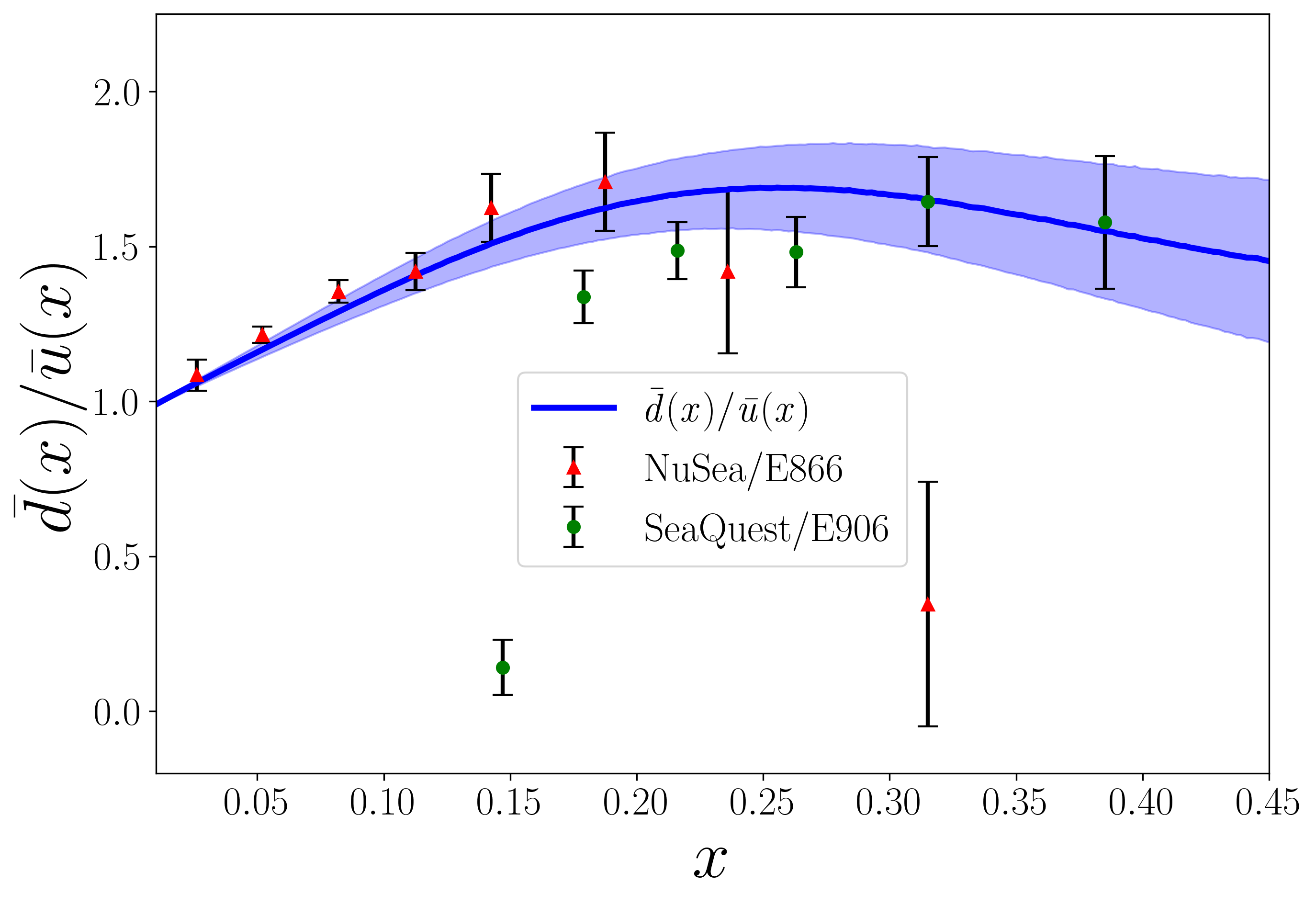}
    \caption{}
  \end{subfigure}
  \caption{The sea quark flavor asymmetry is illustrated in both panels. The left panel (a) shows the difference \(\bar{d}(x) - \bar{u}(x)\), while the right panel (b) presents the ratio \(\bar{d}(x)/\bar{u}(x)\). Experimental data from NuSea/E866 are depicted by red triangles, and data from SeaQuest/E906 are shown as green circles. The theoretical predictions from our model are represented by the blue curve.}
  \label{fig2}
\end{figure}

Using the model wave functions described earlier, we explored the flavor asymmetry in the sea quark distributions to further validate our framework. In particular, we investigated the difference and ratio of the anti-down and anti-up quark distributions, \(\bar{d}(x) - \bar{u}(x)\) and \(\bar{d}(x)/\bar{u}(x)\), respectively. Our predictions are compared with experimental data from the SeaQuest/E906~\citep{dove2023measurement} and NuSea/E866~\citep{towell2001improved} Collaborations, as shown in Fig.~\ref{fig2}. In the region \(x > 0.01\), shown in Fig.~\ref{fig2}(a), our results exhibit good consistency with the available experimental measurements~\citep{dove2023measurement, towell2001improved}. A numerical evaluation of the integral \(\int dx\, [\bar{d}(x) - \bar{u}(x)]\) is provided in Table~\ref{table3}, where our predictions are compared with those from different experiments and model calculations. In the range \(x \in [0.015, 0.35]\), our model predicts
\(\int_{0.015}^{0.35} dx\, [\bar{d}(x) - \bar{u}(x)] = 0.075 \pm 0.009,
\) which is in reasonable agreement with the NuSea/E866 result~\citep{towell2001improved} and the scalar diquark model prediction~\citep{Choudhary:2023unw}. For the narrower range \(x \in [0.13, 0.45]\), our model yields \(\int_{0.13}^{0.45} dx\, [\bar{d}(x) - \bar{u}(x)] = 0.024 \pm 0.0037,\) while the SeaQuest/E906 measurement gives \(0.015 \pm 0.003\)~\citep{dove2023measurement}, which aligns more closely with the scalar diquark model prediction of \(0.015 \pm 0.004\)~\citep{Choudhary:2023unw}. Furthermore, we examined the second moment of the flavor asymmetry:
\(\int_{0.13}^{0.45} dx\, x[\bar{d}(x) - \bar{u}(x)] = 0.005 \pm 0.00085,
\) compared to the SeaQuest/E906 result of \(0.00318^{+0.0005}_{-0.0006}\)~\citep{dove2023measurement} and the scalar diquark model prediction of \(0.003 \pm 0.001\)~\citep{Choudhary:2023unw}.

\begin{table}[hbt]
\caption{The value of the integral \(\int dx\, [\bar{d}(x) - \bar{u}(x)]\) is compared across various experimental results and theoretical models.}
\label{table3}
\setlength{\tabcolsep}{10pt}
\begin{tabular}{lll}
\toprule
\textbf{Model/Experiments} & \textbf{\(x\)-range}  & \textbf{\(\int dx [\bar{d}(x)-\bar{u}(x)]\)} \\
\midrule

\textbf{This work} & \(0.015-0.35\) & \(0.075\pm0.009\) \\

\textbf{Scalar diquark model \citep{Choudhary:2023unw}} & \(0.015-0.35\)  & \(0.069\pm0.015\) \\

\textbf{NuSea/E866 \citep{towell2001improved}} & \(0.015-0.35\)  &  \( 0.0803 \pm 0.011\)  \\

\textbf{SeaQuest/E906 \citep{dove2023measurement}} & \(0.13-0.45\)  & \(0.015 \pm 0.003\) \\

\textbf{This work} & \(0.13-0.45\) & \(0.024\pm0.0037\) \\

\textbf{Scalar diquark model \citep{Choudhary:2023unw}} & \(0.13-0.45\)  & \(0.015 \pm 0.004\) \\
\bottomrule
\end{tabular}
\end{table}

In Fig.~\ref{fig2}(b), we present our results for the ratio \(\bar{d}(x)/\bar{u}(x)\) and compare them with the experimental data from the NuSea/E866~\citep{towell2001improved} and SeaQuest/
E906~\citep{dove2023measurement} Collaborations. A clear qualitative difference is observed between the two experimental data sets: the NuSea/E866 measurement exhibits a rapid decline of \(\bar{d}(x)/\bar{u}(x)\) for \(x > 0.2\), whereas the SeaQuest/E906 data suggest a flatter behavior in this region. Our model prediction shows better agreement with the SeaQuest/E906 data in the range \(0.015 < x < 0.35\), as evident from Fig.~\ref{fig2}(b). In contrast, the rapid fall observed in the NuSea/E866 data at higher \(x\)-values deviates significantly from our prediction. We also compare our result for the ratio \(\bar{d}(x)/\bar{u}(x)\) at \(x = 0.18\) with values reported by the NA51 Collaboration~\citep{NA51:1994xrz} at CERN, based on Drell–Yan measurements, and with predictions from the scalar diquark model~\citep{Choudhary:2023unw}. The NA51 Collaboration reports
\(\left.\frac{\bar{d}(x)}{\bar{u}(x)}\right|_{x = 0.18} = 1.96 \pm 0.15 \pm 0.19,\) while the scalar diquark model yields \(1.30 \pm 0.08\). Our model predicts \( \left.\frac{\bar{d}(x)}{\bar{u}(x)}\right|_{x = 0.18} = 1.61 \pm 0.096,\) which lies between the NA51 experimental value and the scalar diquark model prediction, indicating a reasonable consistency with both.

\section{CHIRAL EVEN GPDS IN OVERLAP REPRESENTATION}
\label{Sec:III}

The Generalized Parton Distributions (GPDs) are defined as non-forward matrix elements of the quark-quark correlator function. In light cone coordinate, the quark-quark correlator \cite{meissner2009generalized} is commonly written as, 

\begin{align}
    F^{\Gamma}_{[\Lambda^{\prime},\Lambda]}(x,\xi,t)&=\frac{1}{2}\int \frac{dz^-}{2 \pi} e^{i x \bar{P}^+ z^-} \langle p^{\prime},\Lambda^{\prime}|\bar{\psi}\bigl(-\frac{z}{2}\bigl)\Gamma
    \mathcal{W}\bigg(\frac{-z}{2},\frac{z}{2}\bigg)\notag\\
    &\hspace{6cm}{\psi}\bigl(+\frac{z}{2}\bigl)| p,\Lambda\rangle {\biggl |}_{z^+=0,\boldsymbol{z}_T=0}
\label{eq1}    
\end{align}

Where $\Gamma$ is the Dirac bilinear and $\Lambda,\Lambda^{\prime}$ are the target initial and final state helicities. $\mathcal{W}$ is the Wilson line, which plays a critical role in ensuring gauge invariance of the nonlocal operator product. It is a path-ordered exponential that connects the quark fields along the light cone, compensating for gauge transformations along the path and maintaining the physical observability of the GPDs. The general definition of $\mathcal{W}$ is,

\begin{equation}
    \mathcal{W}\bigg(\frac{-z}{2},\frac{z}{2}\bigg)=P \ exp\bigg(i g \int_a^b dx^- A^+(x^- n_-) \bigg)
\end{equation}

Here $P$ denotes the path ordering from $a$ to $b$. We have to consider the light-cone gauge where $A^+=0$. So the value of $\mathcal{W}$ is $1$, which also makes it $SU(3)$ color gauge invariant. 

The key Dirac bilinears relevant to twist-3 chiral-even generalized parton distributions are given by $\Gamma = \gamma^j, \gamma^j \gamma_5$. Consequently, the parameterization of Equation (\ref{eq1}), as provided in \cite{meissner2009generalized, rajan2018lorentz}, for these bilinears is expressed as follows:

\begin{align}
    F^{\gamma^j}_{\Lambda \Lambda'} 
    &= \left[\frac{\Delta^j}{2P^+}(E_{2T}+2\Tilde{H}_{2T}) 
    + \frac{i \Lambda \epsilon^{jk} \Delta^k}{2P^+}(\Tilde{E}_{2T} - \xi E_{2T})\right] \delta_{\Lambda \Lambda'} \notag\\
    &\hspace{1cm}+ \left[\frac{-M(\Lambda \delta_{j1} + i \delta_{j2})}{P^+} H_{2T} 
    - \frac{(\Lambda \Delta^1 + i \Delta^2)\Delta^j}{2M P^+} \Tilde{H}_{2T} \right] \delta_{\Lambda - \Lambda'} \label{eqnpara}
\end{align}

\begin{align}
    F^{\gamma^j \gamma_5}_{\Lambda \Lambda'}
    &=\left[\frac{i \epsilon^{jk} \Delta^k}{2P^+}(E^{'}_{2T} + 2 \Tilde{H}^{'}_{2T}) 
    - \frac{ \Lambda \Delta^j}{2P^+}(\Tilde{E}^{'}_{2T} - \xi E^{'}_{2T})\right] \delta_{\Lambda \Lambda'}\notag\\ 
    &\hspace{1cm}+ \left[\frac{M(\delta_{j1} + i \Lambda \delta_{j2})}{P^+} H^{'}_{2T} 
    - \frac{i \epsilon^{jk}(\Lambda \Delta^1 + i \Delta^2) \Delta^k}{2M P^+} \Tilde{H}^{'}_{2T}\right] \delta_{\Lambda - \Lambda'}\label{eqnpara2}
\end{align}

The antisymmetric tensor is defined by $\epsilon^{ij}=1=-\epsilon^{ji}$ and $P=(p+p^{\prime})/2$ is the average proton momentum, $\Delta=p-p^{\prime}$ is the momentum transfer with $t=\Delta^2=-\Delta_T^2$ and $\xi=-\Delta^+/2P^+$ is the skewness parameter. We denote `$\pm$' as the proton helicity. Since $j=1,2$, the parameterization given in equations (\ref{eqnpara}) and (\ref{eqnpara2}) can be reduced according to possible helicity combinations. These combinations are listed in Appendix \ref{appendix} from equations (\ref{eqn5})-(\ref{eqn12}). Using these equations, we can derive the explicit formula of the GPDs in overlap representations \citep{Jain_2024}.  

For matrix structure \(\Gamma=\gamma^j\), we have

\begin{align}
    -\frac{2i \Delta^1 \Delta^2}{M P^+} \tilde{H}_{2T} &=\big(F_{[+-]}^{\gamma^1}+F_{[-+]}^{\gamma^1}\big)
    -i \big(F_{[+-]}^{\gamma^2}-F_{[-+]}^{\gamma^2}\big)\label{eqnfirst}   \\
    \frac{4 M}{P^+}\bigg( H_{2T}+\frac{\Delta^2_T}{4M^2}\tilde{H}_{2T}\bigg) &=\big(F_{[+-]}^{\gamma^1}-F_{[-+]}^{\gamma^1}\big)
    +i \big(F_{[+-]}^{\gamma^2}+F_{[-+]}^{\gamma^2}\big)\\
    \frac{i\boldsymbol{\Delta}_T^2}{P^+}(\tilde{E}_{2T}-\xi E_{2T})&=\Delta_2\big(F_{[++]}^{\gamma^1}-F_{[--]}^{\gamma^1}\big)
    -\Delta_1 \big(F_{[++]}^{\gamma^2}-F_{[--]}^{\gamma^2}\big)\\
   \frac{\boldsymbol{\Delta}_T^2}{P^+}(E_{2T}+2 \tilde{H}_{2T})&=\Delta_1\big(F_{[++]}^{\gamma^1}+F_{[--]}^{\gamma^1}\big)
    +\Delta_2 \big(F_{[++]}^{\gamma^2}+F_{[--]}^{\gamma^2}\big)
\end{align}    

Similarly, for matrix structure \(\Gamma=\gamma^j\gamma_5\) the following expression can be derived:

\begin{align}    
    \frac{2 i \Delta^1 \Delta^2}{M P^+}\tilde{H}{}_{2T}^{\prime}&=\big(F^{\gamma^1 \gamma_5}_{[+-]}-F^{\gamma^1 \gamma_5}_{[-+]}\big)
    -i \big(F^{\gamma^2 \gamma_5}_{[+-]}+F^{\gamma^2 \gamma_5}_{[-+]}\big)\\
    \frac{i \boldsymbol{\Delta}_T^2}{P^+}(E_{2T}^{\prime}+2 \tilde{H}_{2T}^{\prime})&=\Delta_2\big(F^{\gamma^1 \gamma_5}_{[++]}+F^{\gamma^1 \gamma_5}_{[--]}\big)
    - \Delta_1 \big(F^{\gamma^2 \gamma_5}_{[++]}+F^{\gamma^2 \gamma_5}_{[--]}\big)\\
    \frac{4M}{P^+} \bigg( H_{2T}^{\prime}+\frac{\boldsymbol{\Delta}_T^2}{4 M^2} \tilde{H}_{2T}^{\prime}\bigg)&=\big(F^{\gamma^1 \gamma_5}_{[+-]}+F^{\gamma^1 \gamma_5}_{[-+]}\big)
    +i \big(F^{\gamma^2 \gamma_5}_{[+-]}-F^{\gamma^2 \gamma_5}_{[-+]}\big)\\
    \frac{\boldsymbol{\Delta}_T^2}{P^+}(\tilde{E}_{2T}^{\prime}-\xi E_{2T}^{\prime})&=\Delta_1 \big(F^{\gamma^1 \gamma_5}_{[++]}-F^{\gamma^1 \gamma_5}_{[--]}\big)
    +\Delta_2 \big(F^{\gamma^2 \gamma_5}_{[++]}-F^{\gamma^2 \gamma_5}_{[--]}\big)\label{eqnlast}
\end{align}

The quark-quark correlator, as presented in \citep{sharma2024exploring}, is expressed in the overlap representation at $\xi=0$ as,

\begin{align}
    F^\Gamma_{[\Lambda^{\prime}\Lambda]}(x,0,t)
    &=\int \frac{d^2 \boldsymbol{k}_T}{16 \pi^3} \sum_{\lambda_{q_i}} \sum_{\lambda_{q_f}}\psi^{\Lambda^{\prime}\dagger}_{\lambda_{q_f}\lambda_A}(x,\boldsymbol{k}_T^{\prime \prime})\psi^{\Lambda}_{\lambda_{q_i}\lambda_A}(x,\boldsymbol{k}_T^{\prime})\nonumber\\
    &\hspace{2cm}\frac{u^{\dagger}_{\lambda_{q_f}}(x P^+,\boldsymbol{k}_T+\frac{\boldsymbol{\Delta}_T}{2})\gamma^0 \Gamma u_{\lambda_{q_i}}(x P^+,\boldsymbol{k}_T-\frac{\boldsymbol{\Delta}_T}{2})}{2 x P^+}
\end{align}

In this context, \(\lambda_{q_i}\) and \(\lambda_{q_f}\) denote the initial and final helicities of the sea quark, respectively. In the overlap representation, \(\Delta\) is treated as a two-dimensional complex vector, defined by \(\Delta = \Delta^1 + i \Delta^2\). The expression 
\[
u^{\dagger}_{\lambda_{q_f}} \left( x P^+, \boldsymbol{k}_T + \frac{\boldsymbol{\Delta}_T}{2} \right) \gamma^0 \Gamma u_{\lambda_{q_i}} \left( x P^+, \boldsymbol{k}_T - \frac{\boldsymbol{\Delta}_T}{2} \right)
\]
represents the spinor product associated with higher-twist Dirac matrices. The Dirac matrices within the light-cone formalism are comprehensively detailed in \cite{brodsky1998quantum, harindranath1996introduction}. The initial and final transverse momenta of the struck sea quark are,

\begin{align}
    \boldsymbol{k}_T^{\prime \prime}= \boldsymbol{k}_T+(1-x)\frac{\boldsymbol{\Delta}_T}{2}, && \boldsymbol{k}_T^{\prime}= \boldsymbol{k}_T-(1-x)\frac{\boldsymbol{\Delta}_T}{2}
\end{align}

At zero skewness (\(\xi = 0\)), only four chiral GPDs remain, as shown using the symmetry properties of GPDs in \cite{zhang2024twist}. Therefore, the chiral-even GPDs of sea quarks can be derived from equations (\ref{eqnfirst})-(\ref{eqnlast}) as follows,

\begin{align}
    x \tilde{E}_{2T}(x, \boldsymbol{\Delta}_T^2) &= -\int \frac{d^2 \boldsymbol{k}_T}{16 \pi^3} \Bigg\{ \bigg(\frac{x M+m}{x}\bigg)^2\varphi_1^{*}\varphi_1 {-}\bigg[\bigg(\boldsymbol{k}_T^2-\frac{1}{4} (1-x)^2\boldsymbol{\Delta}_T^2 \bigg)- \notag \\ 
    &\hspace{2cm}2\bigg(\frac{\boldsymbol{k}_T^2 \boldsymbol{\Delta}_T^2-(\boldsymbol{k}_T \cdot \boldsymbol{\Delta}_T)^2}{\boldsymbol{\Delta}_T^2}\bigg)(1-x)\bigg]\frac{\varphi_2^{*}\varphi_2 }{x^2}\Bigg\},\label{GPD1}
\end{align}    
\begin{align}    
    x{H}'_{2T}(x, \boldsymbol{\Delta}_T^2)&= -\frac{1}{M }\int \frac{d^2 \boldsymbol{k}_T}{16 \pi^3} \Bigg\{{m} \bigg(\frac{x M+m}{x}\bigg)^2\varphi_1^{*}\varphi_1 + 2 \frac{(\boldsymbol{k}_T \cdot \boldsymbol{\Delta}_T)^2}{\boldsymbol{\Delta}_T^2}\frac{\varphi_1^{*}\varphi_2}{x^2}\notag \\ 
    &-\frac{m}{x^2}\bigg[\bigg(\frac{\boldsymbol{k}_T^2 \boldsymbol{\Delta}_T^2-2 (\boldsymbol{k}_T \cdot \boldsymbol{\Delta}_T)^2}{\boldsymbol{\Delta}_T^2}\bigg)+\frac{1}{4} (1-x)^2\boldsymbol{\Delta}_T^2 \bigg]\varphi_2^{*}\varphi_2 \Bigg\},\label{H2prim2T}
\end{align}
\begin{align}
    &x\tilde{H}'_{2T}(x, \boldsymbol{\Delta}_T^2)\notag\\
    &={-}\frac{M}{\boldsymbol{\Delta}_T^2}\int \frac{d^2 \boldsymbol{k}_T}{4 \pi^3} \Bigg\{(x M+m)\bigg[\bigg(\frac{\boldsymbol{k}_T^2 \boldsymbol{\Delta}_T^2-2 (\boldsymbol{k}_T \cdot \boldsymbol{\Delta}_T)^2}{\boldsymbol{\Delta}_T^2}\bigg)+\frac{1}{4} (1-x)\boldsymbol{\Delta}_T^2 \bigg]\notag \\
    &\hspace{1.5cm}\frac{\varphi_1^{*}\varphi_2}{x^2}-m\bigg[\bigg(\frac{\boldsymbol{k}_T^2 \boldsymbol{\Delta}_T^2-2 (\boldsymbol{k}_T \cdot \boldsymbol{\Delta}_T)^2}{\boldsymbol{\Delta}_T^2}\bigg)+\frac{1}{4} (1-x)^2\boldsymbol{\Delta}_T^2 \bigg]\frac{\varphi_2^{*}\varphi_2}{x^2} \Bigg\},
\end{align}
\begin{align}
    &x{E}'_{2T}(x, \boldsymbol{\Delta}_T^2)\notag\\
    &=\int \frac{d^2 \boldsymbol{k}_T}{4 \pi^3} \Bigg\{-\bigg(\frac{x M+m}{x}\bigg)^2 \varphi_1^{*}\varphi_1+2 m (x M+m)\frac{(1-x)}{x^2}\varphi_1^{*}\varphi_2+\notag \\ 
    &\hspace{1.5cm}\bigg[2\bigg(\frac{\boldsymbol{k}_T^2 \boldsymbol{\Delta}_T^2- (\boldsymbol{k}_T \cdot \boldsymbol{\Delta}_T)^2}{\boldsymbol{\Delta}_T^2}\bigg)-\bigg(\boldsymbol{k}_T^2-\frac{1}{4} (1-x)^2\boldsymbol{\Delta}_T^2\bigg) \bigg]\frac{\varphi_2^{*}\varphi_2}{x^2}+\frac{2M}{\boldsymbol{\Delta}_T^2}\notag \\
    &\hspace{2cm} \Bigg\{(x M+m)\bigg[\bigg(\frac{\boldsymbol{k}_T^2 \boldsymbol{\Delta}_T^2-2 (\boldsymbol{k}_T \cdot \boldsymbol{\Delta}_T)^2}{\boldsymbol{\Delta}_T^2}\bigg)+\frac{1}{4} (1-x)\boldsymbol{\Delta}_T^2 \bigg]\frac{\varphi_1^{*}\varphi_2}{x^2}\notag \\
    &\hspace{2cm}-m\bigg[\bigg(\frac{\boldsymbol{k}_T^2 \boldsymbol{\Delta}_T^2-2 (\boldsymbol{k}_T \cdot \boldsymbol{\Delta}_T)^2}{\boldsymbol{\Delta}_T^2}\bigg)+\frac{1}{4} (1-x)^2\boldsymbol{\Delta}_T^2 \bigg]\frac{\varphi_2^{*}\varphi_2}{x^2} \Bigg\}.\label{GPD4}
\end{align}

\section{Result and Discussion}
\label{Results}

In this section, we present the numerical results for the twist-3 chiral-even generalized parton distributions (GPDs) of light sea quarks in the proton at zero skewness, corresponding to the Dirac bilinear structures \(\gamma^i\) and \(\gamma_i \gamma_5\). We also investigate the twist-3 parton distribution function \(g_T(x)\), which encapsulates quark-gluon correlations beyond leading twist. In addition, we examine the validity of the Burkhardt--Cottingham sum rule~\cite{burkhardt1970sum} separately for the \(\bar{u}\) and \(\bar{d}\) distributions within our model framework. The kinetic OAM is also investigated in the final subsection.

\subsection{Results for chiral even twist-3 GPDs}

We solved chiral even GPDs within our model using the overlap representation presented in equations \eqref{GPD1}-\eqref{GPD4}. Since $\xi=0$, we have investigated the GPDs as a function of $x$ and $\boldsymbol{\Delta}_T$. In our calculation, we have found that $x\tilde{E}_{2T},x\tilde{H}_{2T}^{\prime},xE_{2T}^{\prime}$ and $xH_{2T}^{\prime}$ are non-zero and all other chiral even GPDs are zero, which is consistent with \cite{zhang2024twist}.

\begin{figure}[htbp]
    \centering
    \begin{subfigure}[b]{0.45\textwidth}
        \centering
        \includegraphics[width=\textwidth]{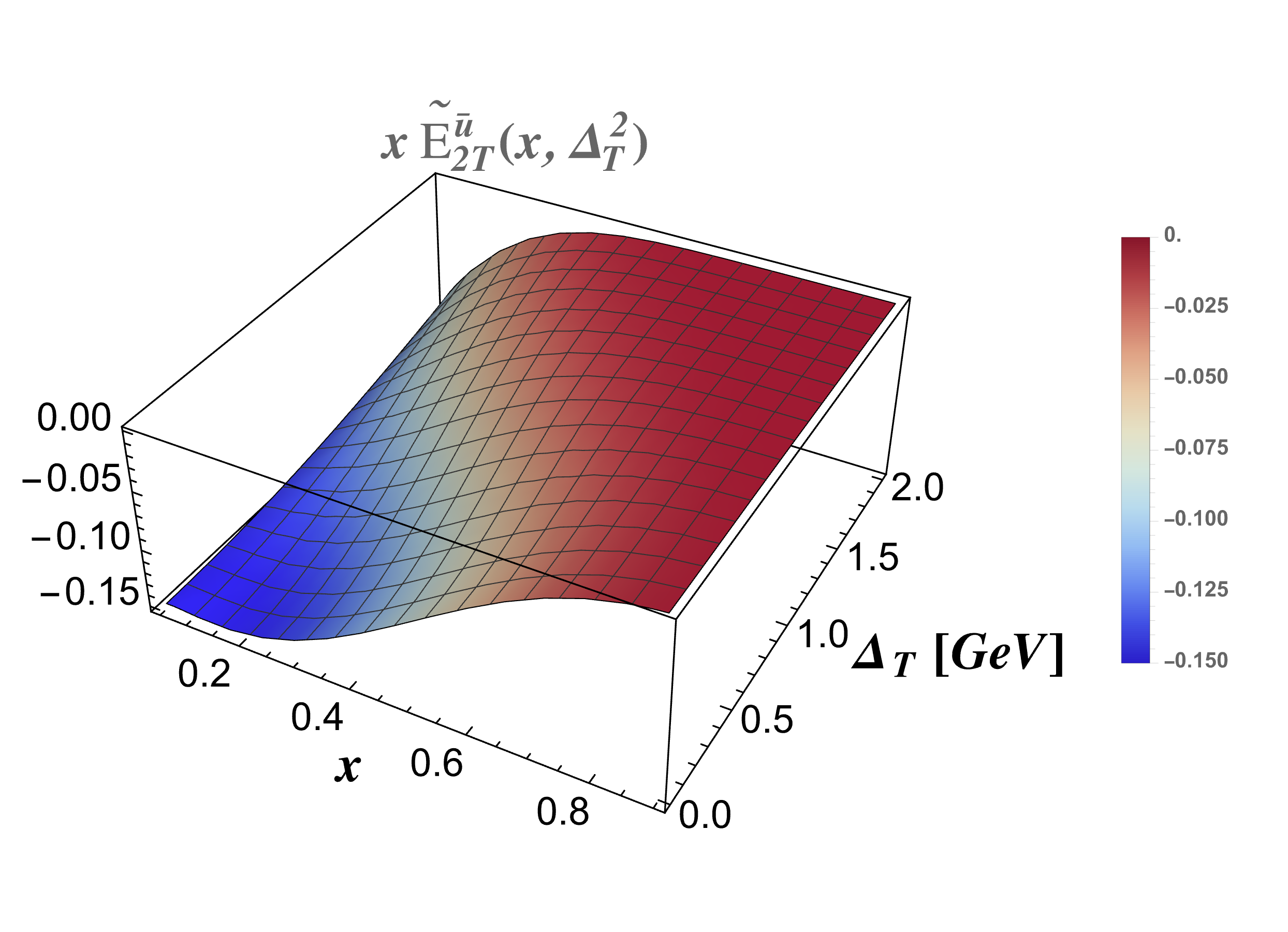} 
        \caption{}
    \end{subfigure}
    \hspace{1cm}
    \begin{subfigure}[b]{0.45\textwidth}
        \centering
        \includegraphics[width=\textwidth]{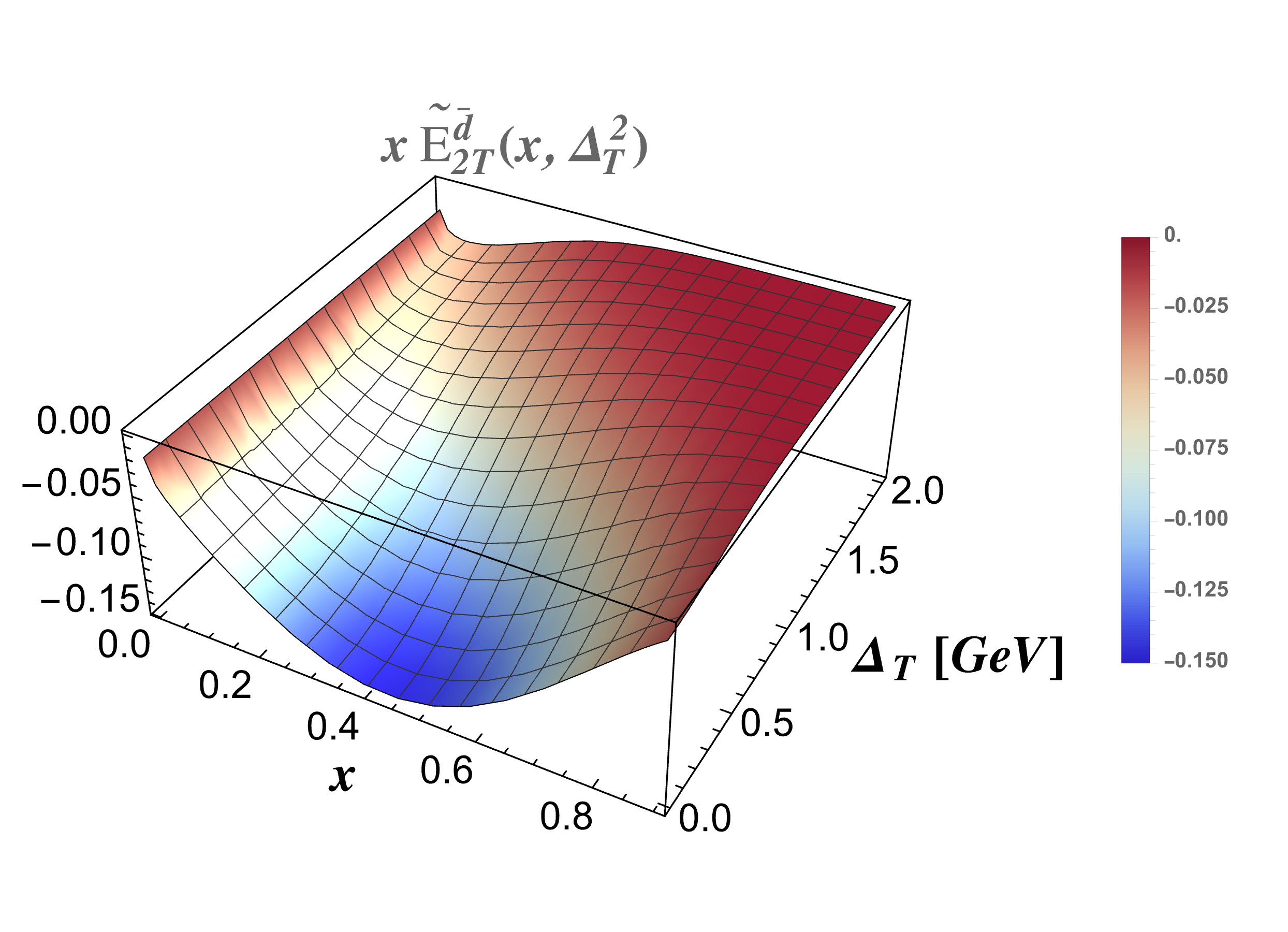} 
        \caption{}
    \end{subfigure}
    \caption{The twist-3 chiral-even GPDs of light sea quarks, corresponding to the Dirac structure \(\Gamma = \gamma^i\), are presented. Panel (a) shows the distribution for the \(\bar{u}\) quark, while panel (b) displays the corresponding distribution for the \(\bar{d}\) quark.}
    \label{Figure3}
\end{figure}

\begin{figure}[htbp]
    \centering
    \begin{subfigure}[b]{0.45\textwidth}
        \centering
        \includegraphics[width=\textwidth]{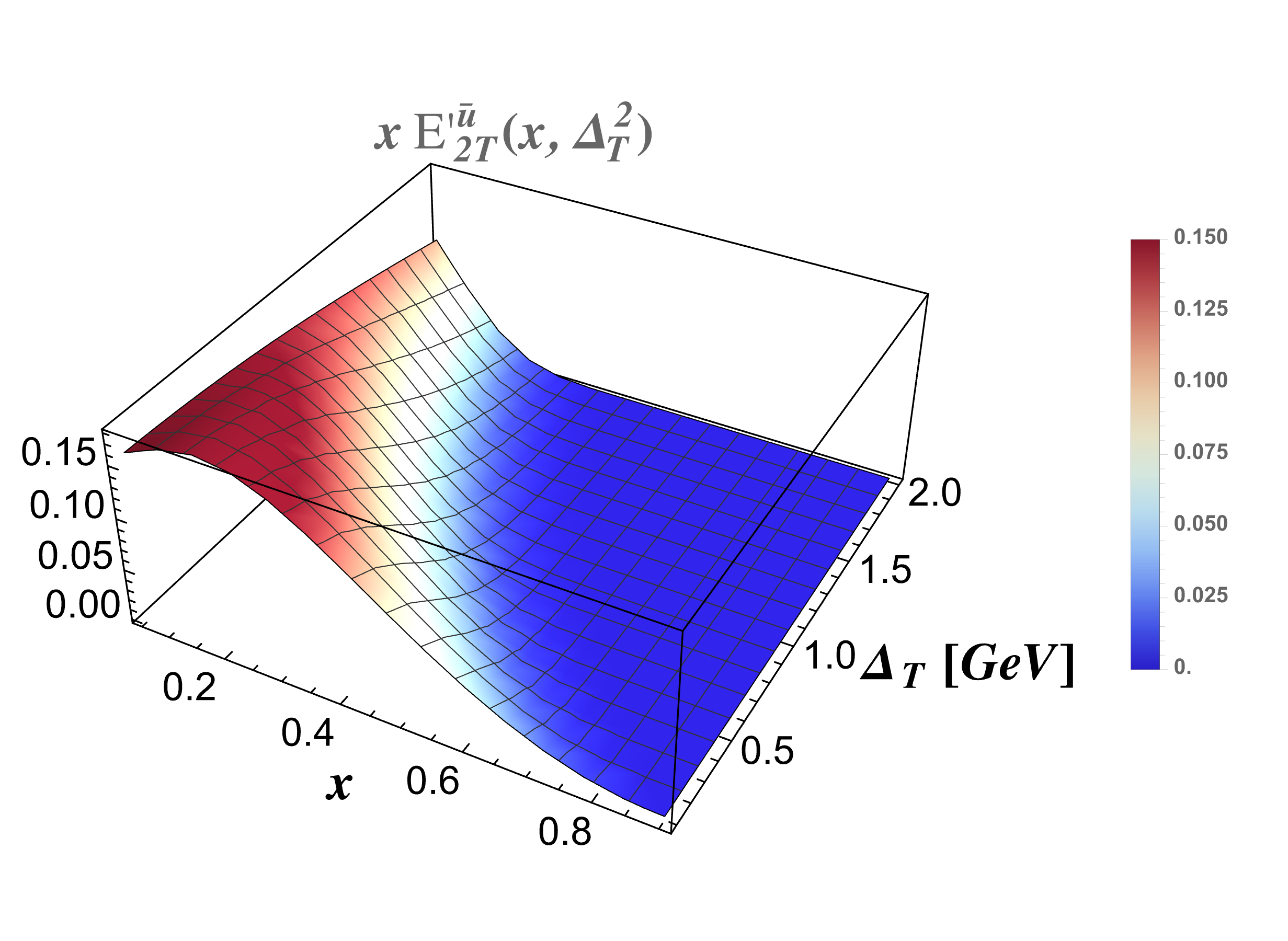} 
        \caption{}
    \end{subfigure}
    \hspace{1cm}
    \begin{subfigure}[b]{0.45\textwidth}
        \centering
        \includegraphics[width=\textwidth]{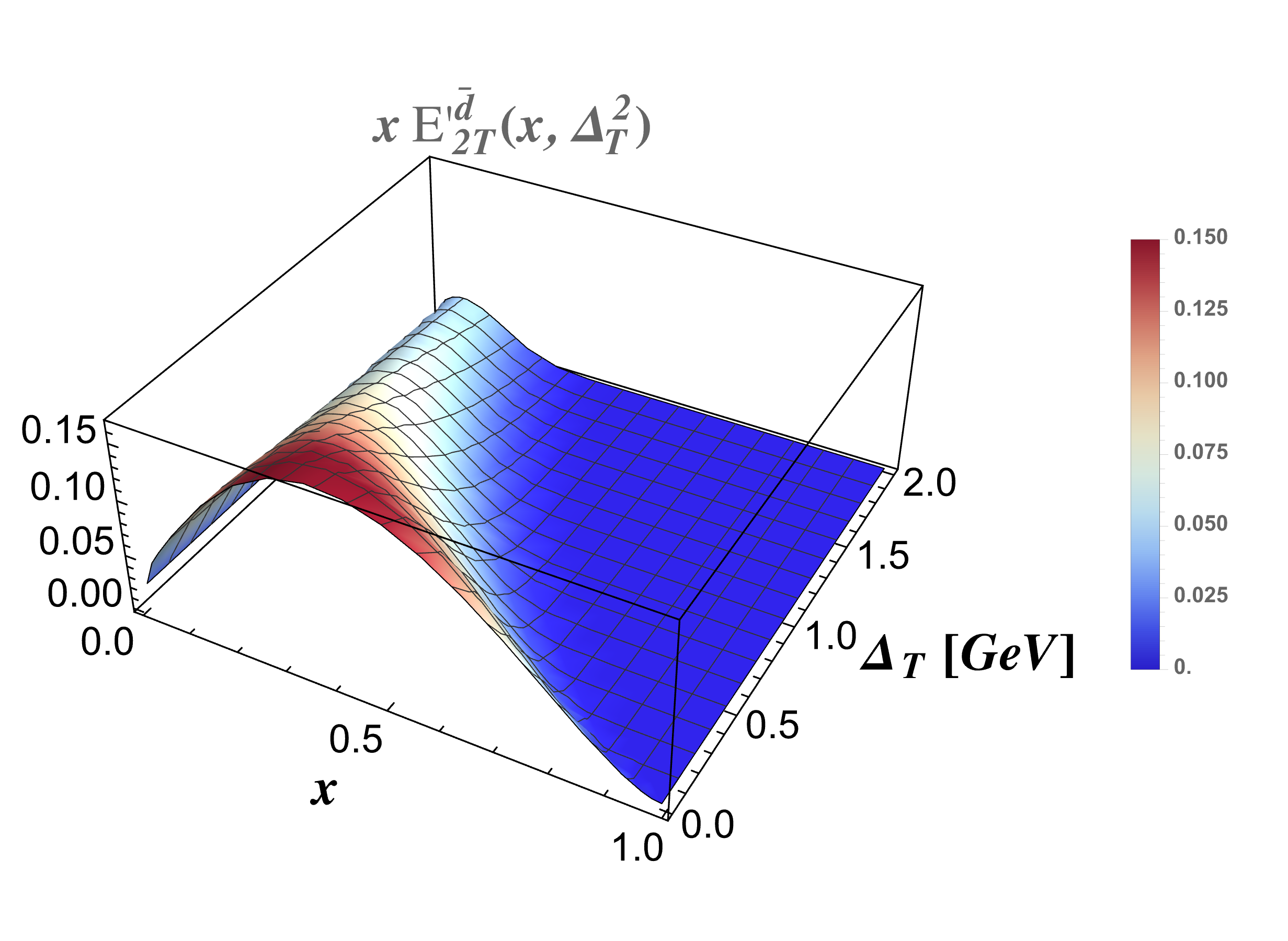} 
        \caption{}
    \end{subfigure}

    \vspace{1em}

    \begin{subfigure}[b]{0.45\textwidth}
        \centering
        \includegraphics[width=\textwidth]{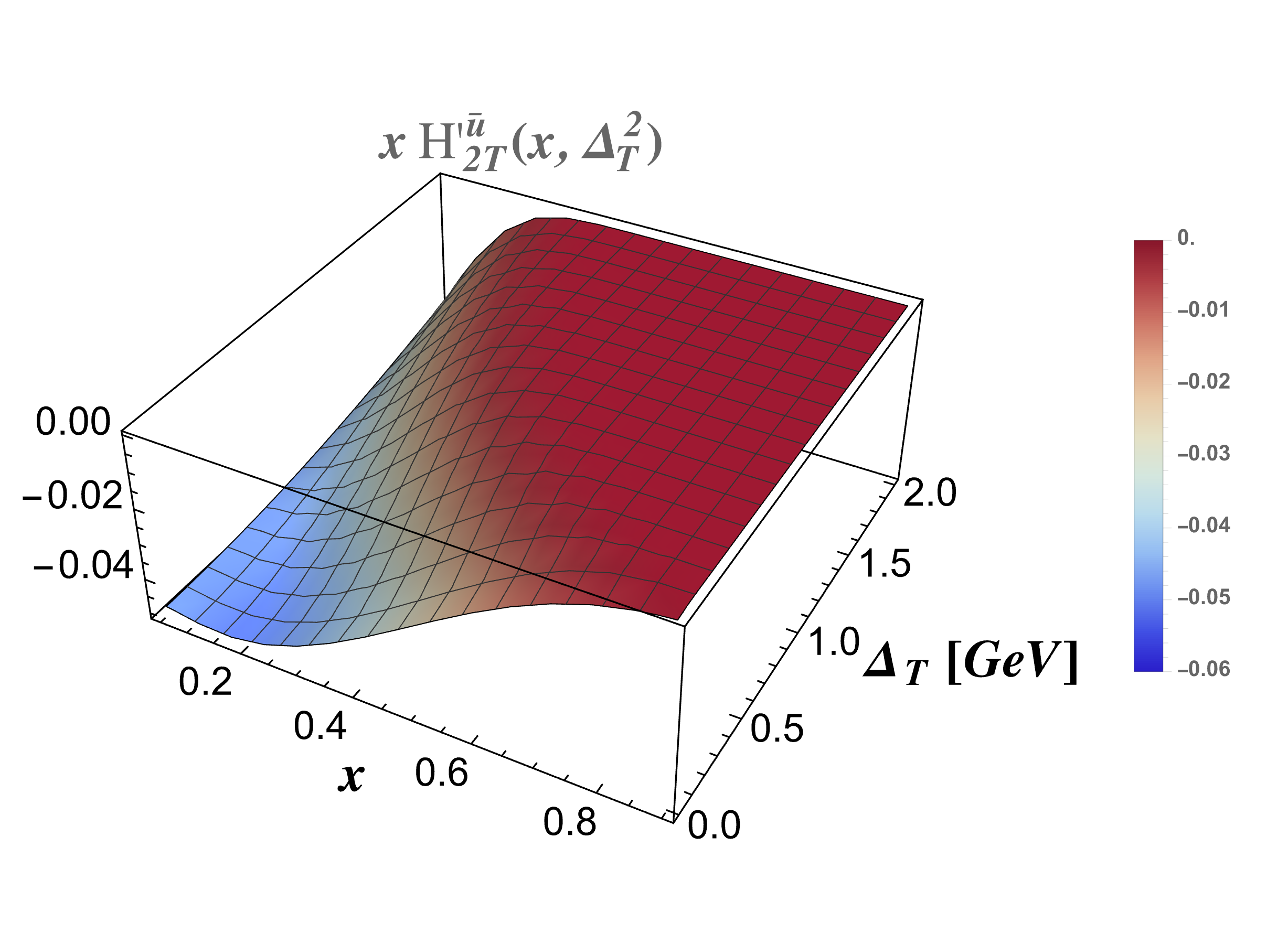} 
        \caption{}
    \end{subfigure}
    \hspace{1cm}
    \begin{subfigure}[b]{0.45\textwidth}
        \centering
        \includegraphics[width=\textwidth]{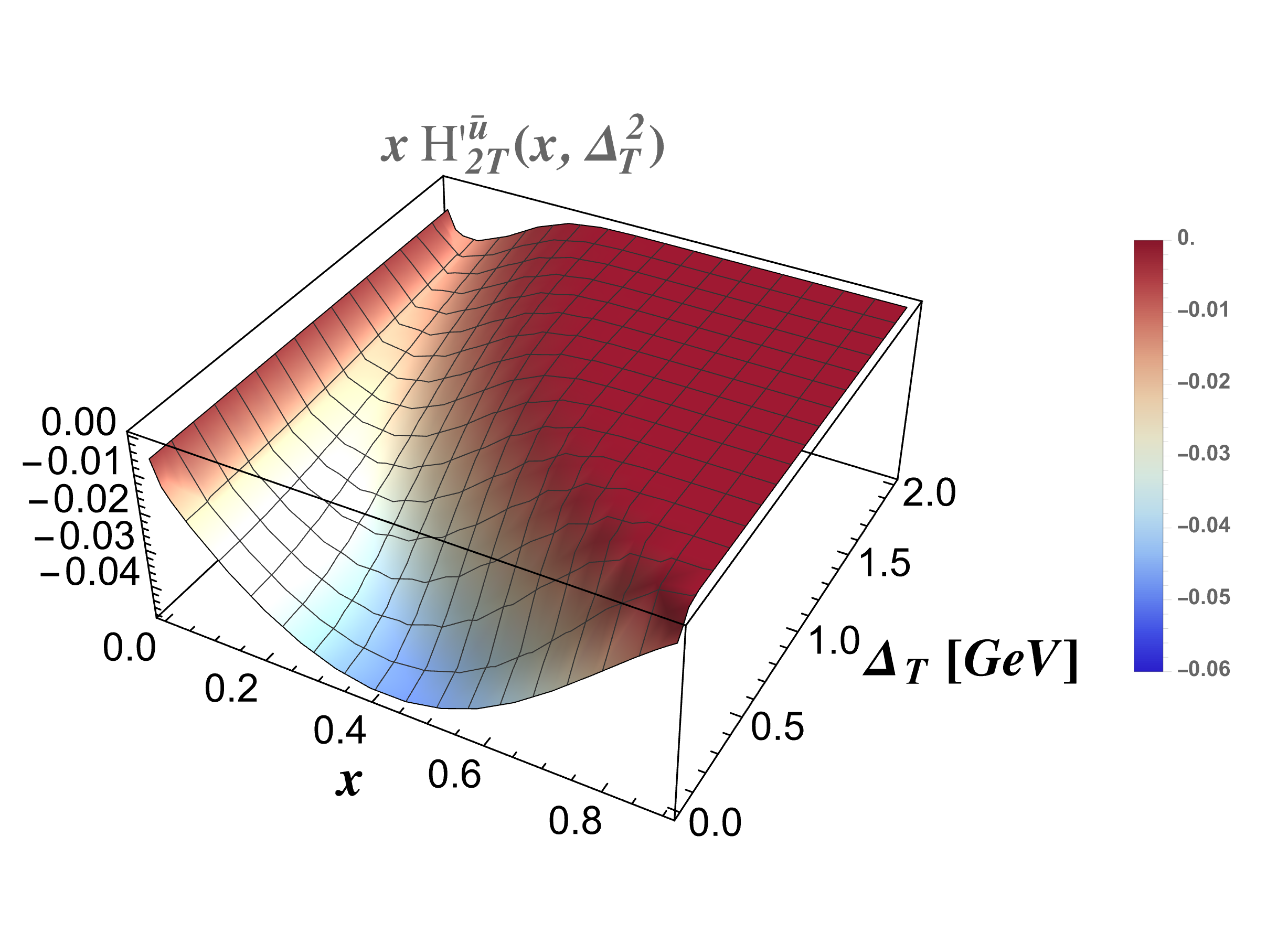} 
        \caption{}
    \end{subfigure}

    \vspace{1em}

    \begin{subfigure}[b]{0.45\textwidth}
        \centering
        \includegraphics[width=\textwidth]{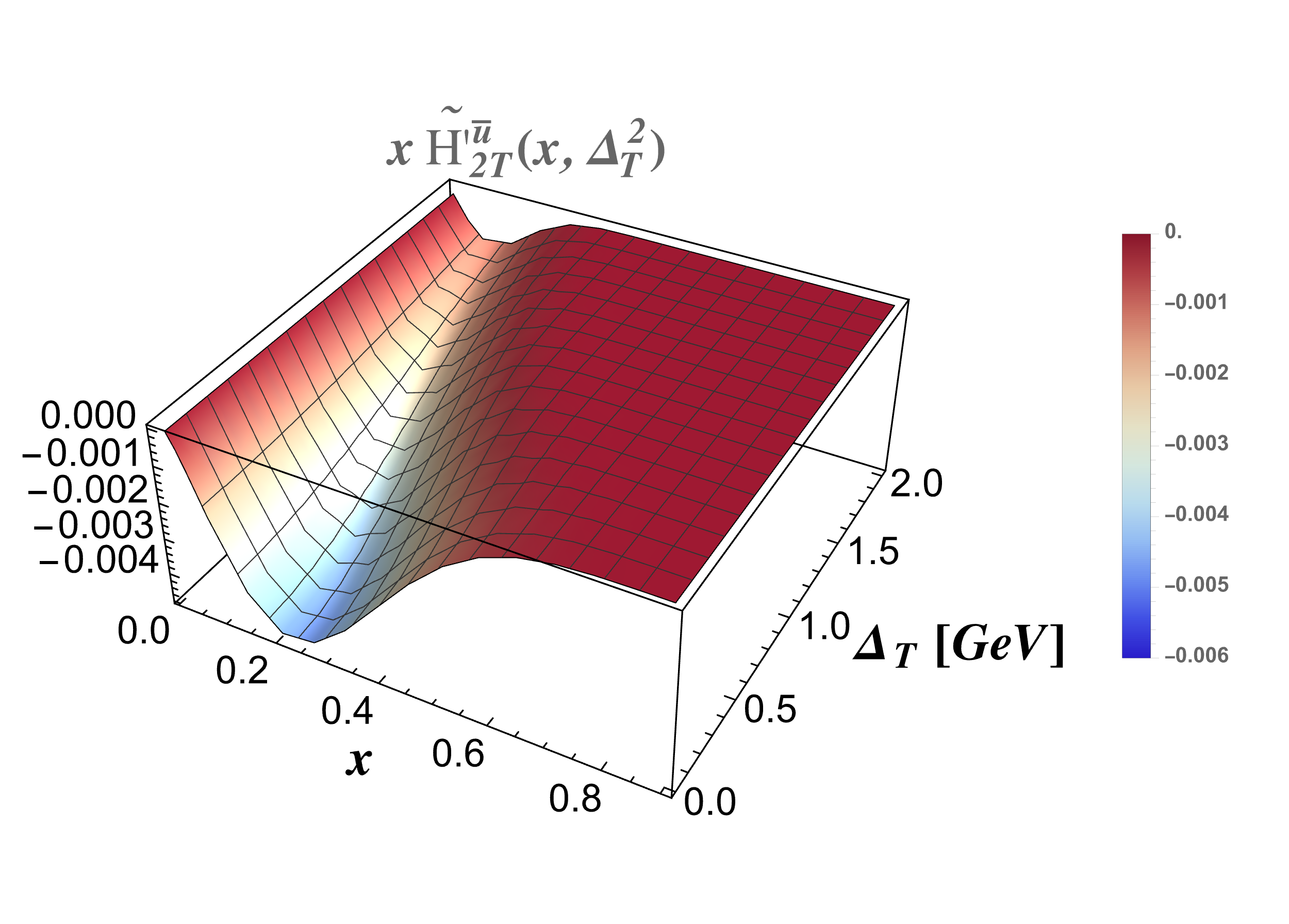} 
        \caption{}
    \end{subfigure}
    \hspace{1cm}
    \begin{subfigure}[b]{0.45\textwidth}
        \centering
        \includegraphics[width=\textwidth]{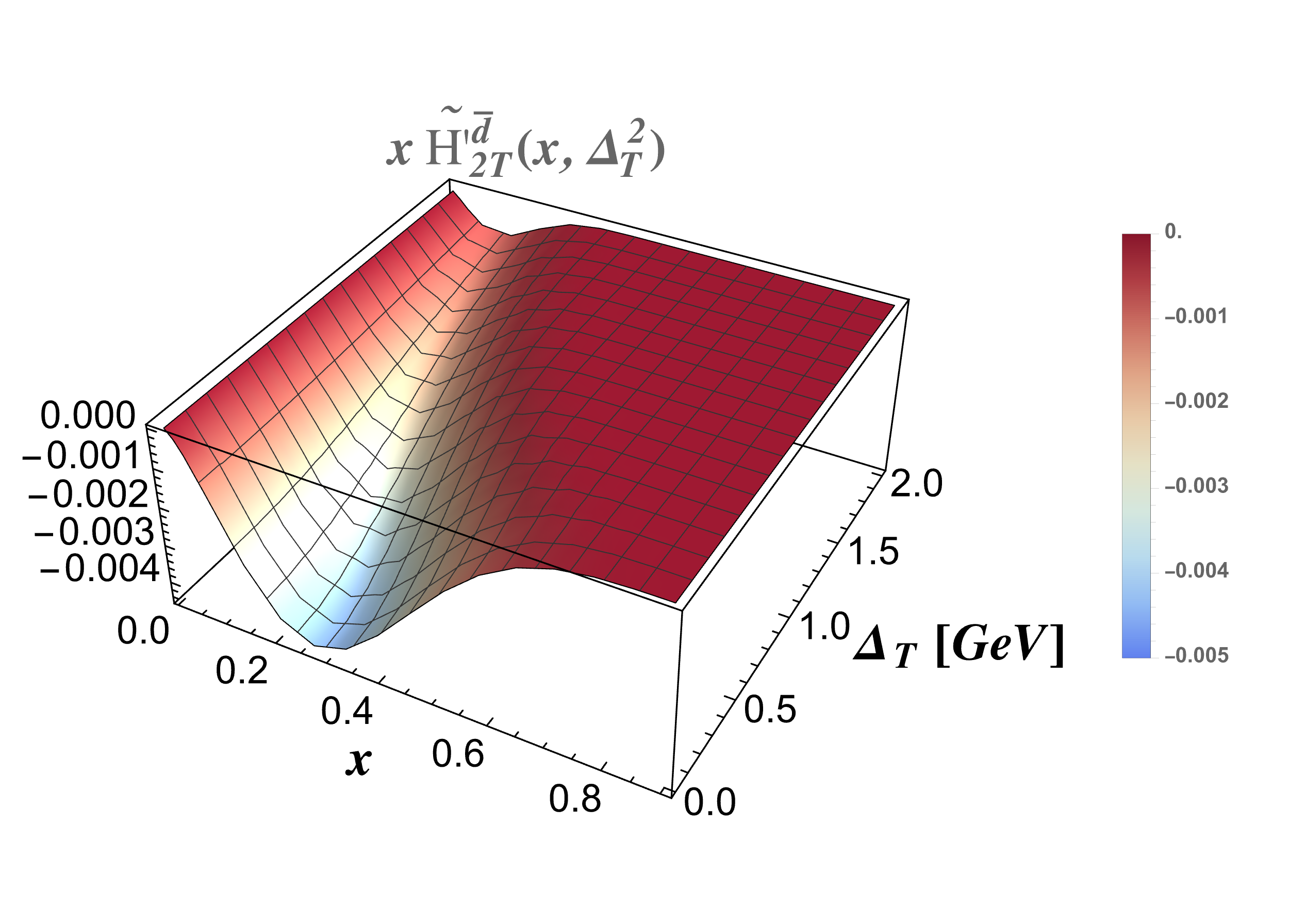} 
        \caption{}
    \end{subfigure}

    \caption{The twist-3 chiral-even GPDs of light sea quarks, corresponding to the Dirac structure \(\Gamma = \gamma^i \gamma_5\), are plotted as functions of the longitudinal momentum fraction \(x\) and transverse momentum transfer \(\boldsymbol{\Delta}_T\) in the range \(0.005 < x < 0.6\) and \(0 < \boldsymbol{\Delta}_T < 2\) GeV.}
    \label{Figure4}
\end{figure}

At zero skewness, the only surviving generalized parton distribution with a 
\(\gamma^i\) Dirac structure is \(x\tilde{E}_{2T}\). In Fig.~\ref{Figure3}, we present the distributions of \(x\tilde{E}_{2T}\) as functions of the longitudinal momentum fraction \(x\) and the transverse momentum transfer \(\boldsymbol{\Delta}_T\) for both \(\bar{u}\) and \(\bar{d}\) sea quarks. Within the kinematic range 
\(0.005 < x < 0.9\) and \(0 < \boldsymbol{\Delta}_T < 2 \, \text{GeV}\), both 
\(x\tilde{E}_{2T}^{\bar{u}}\) and \(x\tilde{E}_{2T}^{\bar{d}}\) are found to be negative at small \(x\) and low transverse momentum transfer. The peak of the distribution appears at small \(x\) and low \(\boldsymbol{\Delta}_T\) for the \(\bar{u}\) distribution, whereas the \(\bar{d}\) distribution exhibits a peak around \(x \approx 0.45\).  
A similar behavior is observed for the GPDs associated with the 
\(\gamma^i\gamma_5\) Dirac structure.  In Fig.~\ref{Figure4}, we display the distributions of \(xE_{2T}^{\prime}\), 
\(xH_{2T}^{\prime}\), and \(x\tilde{H}_{2T}^{\prime}\) as functions of \(x\) and 
\(\boldsymbol{\Delta}_T\). The GPD \(xH_{2T}^{\prime}\) remains negative throughout the entire kinematic region \(0.005 < x < 0.9\) and \(0 < \boldsymbol{\Delta}_T < 2 \, \text{GeV}\).  The GPD \(xE_{2T}^{\prime}\) is positive at small \(x\) and low 
\(\boldsymbol{\Delta}_T\), while \(x\tilde{H}_{2T}^{\prime}\) remains negative across the full kinematic range and approaches zero as \(x \to 0\). These differences in peak structures between the \(\bar{u}\) and \(\bar{d}\) quark distributions further illustrate the effects of sea quark flavor asymmetry. We compared our distributions with those discussed in the Light-Cone Model (Twist-3)~\cite{34lm-pr91}. Based on the descriptions in that work, we find that the signs of the GPDs are in good agreement, while the magnitudes predicted by our model are systematically smaller. No direct numerical or graphical comparison was performed due to differences in model frameworks and parameter choices.

\begin{figure}[htbp]
    \centering
    \begin{subfigure}[b]{0.4\textwidth}
        \centering
        \includegraphics[width=\textwidth]
        {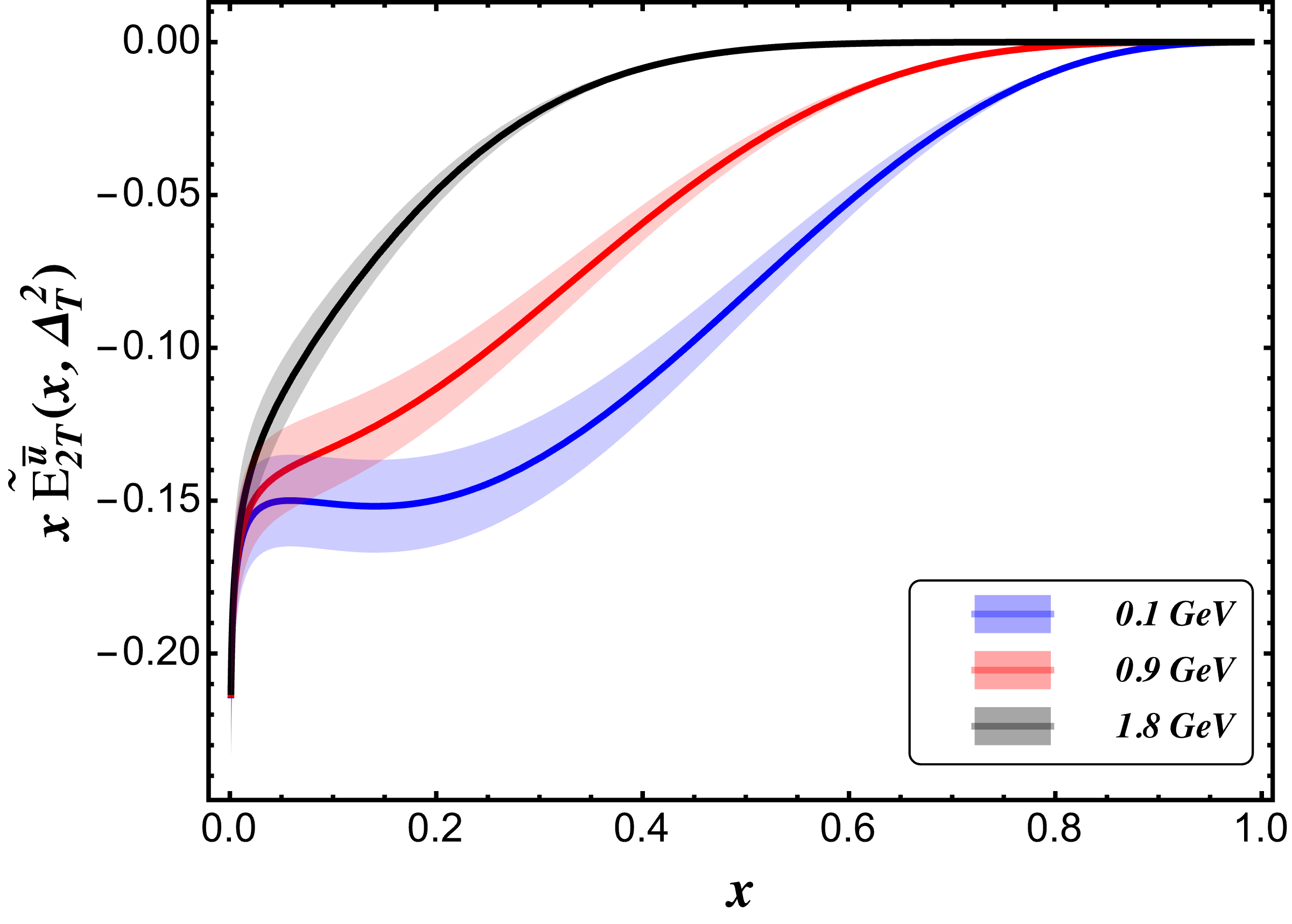}
        \caption{}
    \end{subfigure}
    \hspace{1cm}
    \begin{subfigure}[b]{0.4\textwidth}
        \centering
        \includegraphics[width=\textwidth]{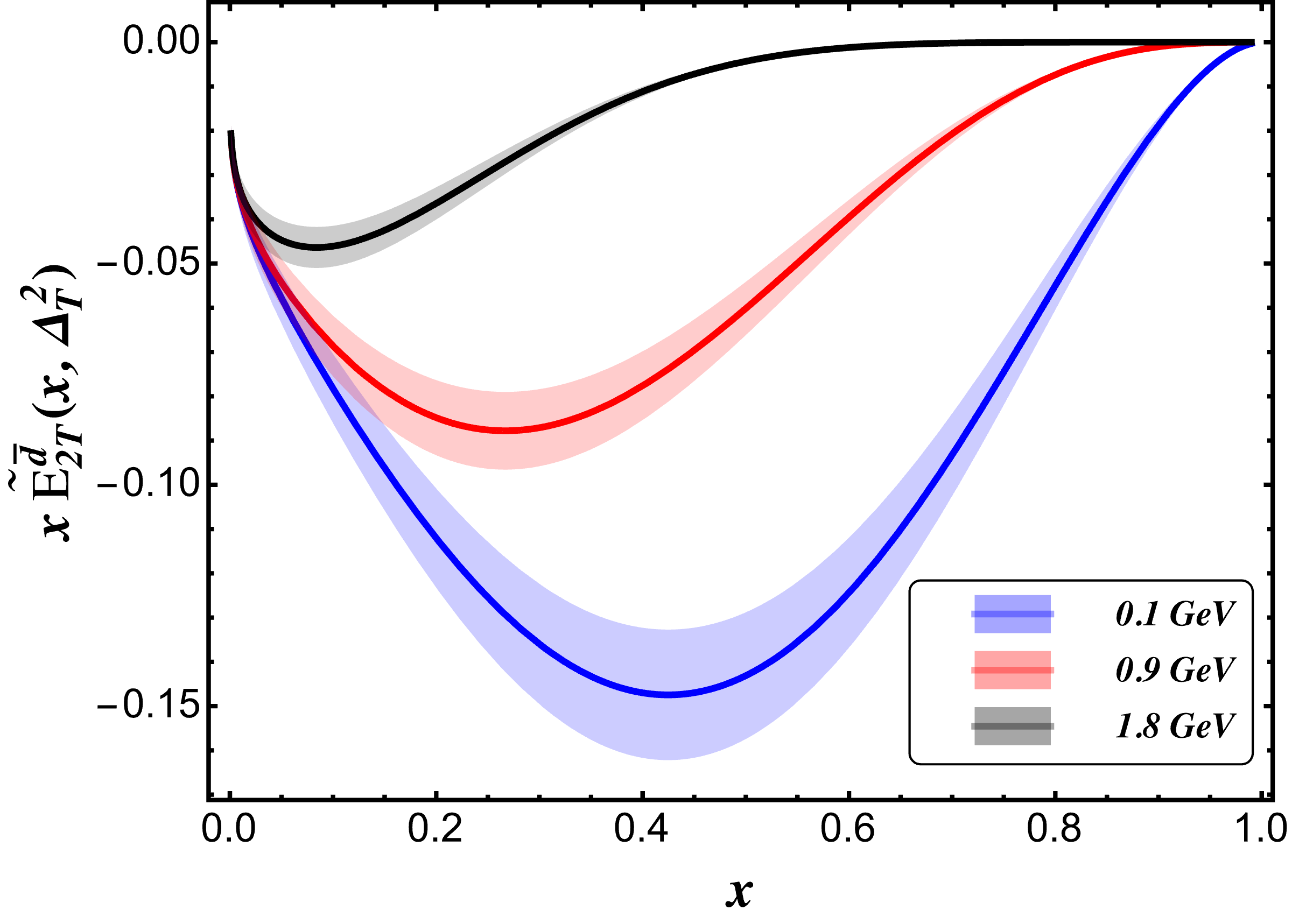} 
        \caption{}
    \end{subfigure}
    \caption{The twist-3 chiral-even generalized parton distribution (GPD) \(x \tilde{E}_{2T}\), corresponding to the \(\gamma^i\) Dirac structure, is plotted as a function of \(x\) for fixed values of the transverse momentum transfer: \(\boldsymbol{\Delta}_T = 0.1 \ \text{GeV}\) (blue), \(0.9 \ \text{GeV}\) (red), and \(1.8 \ \text{GeV}\) (black).}
    \label{figure5}
\end{figure}

To gain a deeper insight into the structure of the chiral-even twist-3 generalized parton distributions (GPDs), we present two-dimensional plots of \( x\tilde{E}_{2T} \) at fixed values of the transverse momentum transfer \(\boldsymbol{\Delta}_T\). Figure~\ref{figure5} displays the variation of \( x\tilde{E}_{2T} \) as a function of the longitudinal momentum fraction \( x \), for representative values \(\boldsymbol{\Delta}_T = 0.1\), \(0.9\), and \(1.8\, \text{GeV}\). It is evident that the magnitude of \( x\tilde{E}_{2T} \) increases with increasing \(\boldsymbol{\Delta}_T\) and becomes zero at higher values of \(\boldsymbol{\Delta}_T\).

\begin{figure}[htbp]
    \centering   
    \begin{subfigure}[b]{0.4\textwidth}
        \centering
        \includegraphics[width=\textwidth]{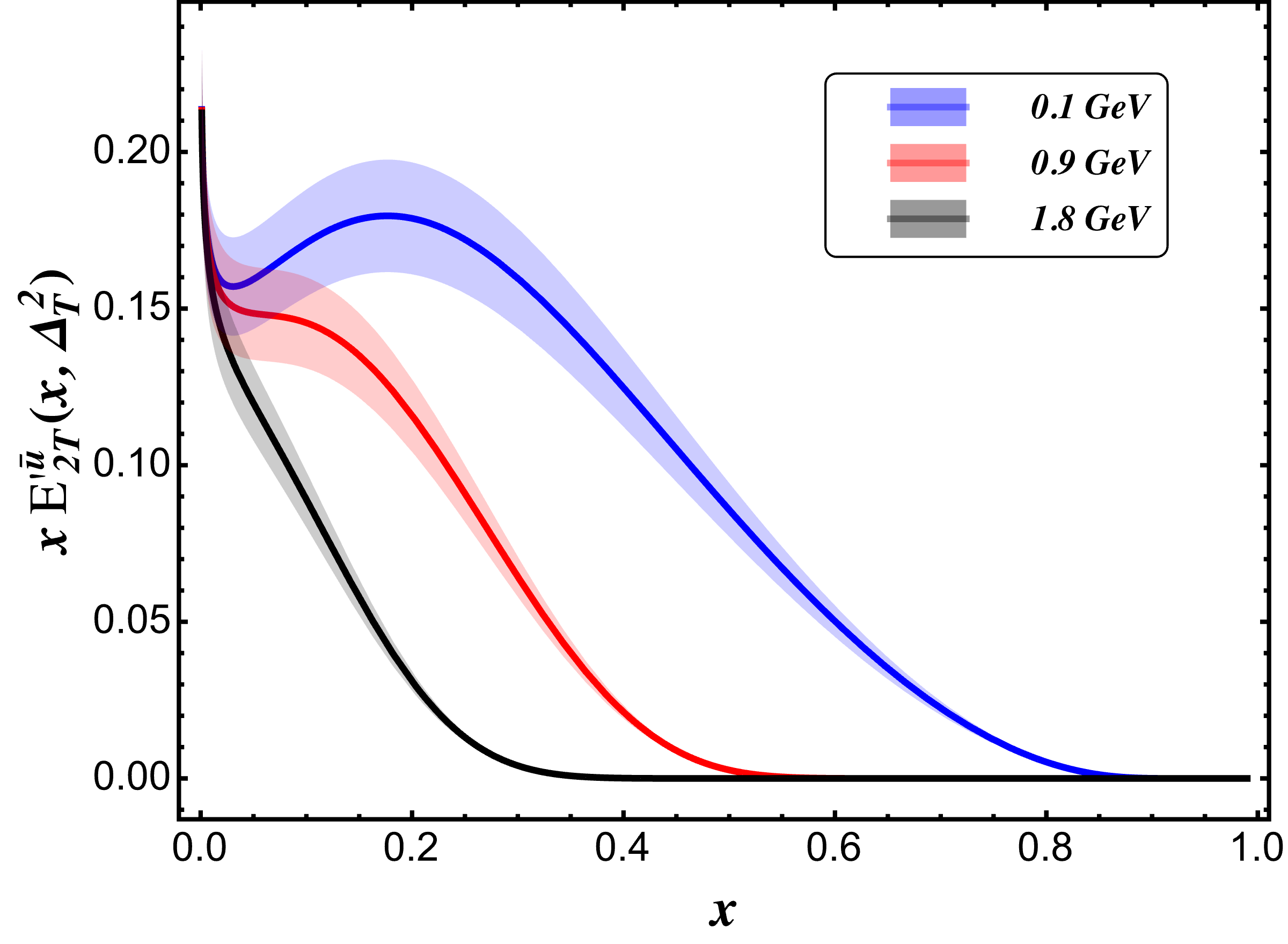} 
        \caption{}
    \end{subfigure}
    \hspace{1cm}
    \begin{subfigure}[b]{0.4\textwidth}
        \centering
        \includegraphics[width=\textwidth]{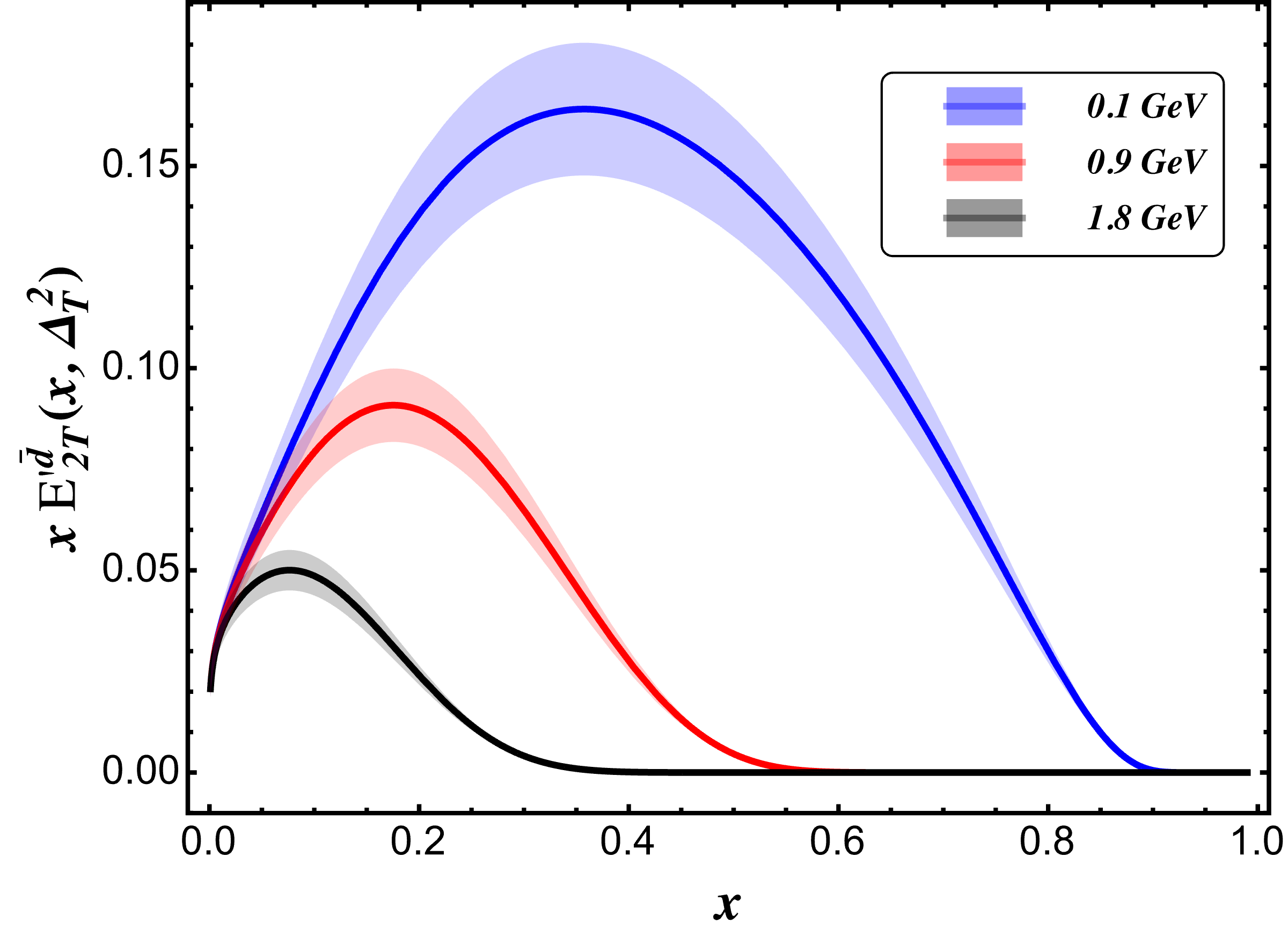} 
        \caption{}
    \end{subfigure}

    \vspace{1em}

    \begin{subfigure}[b]{0.4\textwidth}
        \centering
        \includegraphics[width=\textwidth]{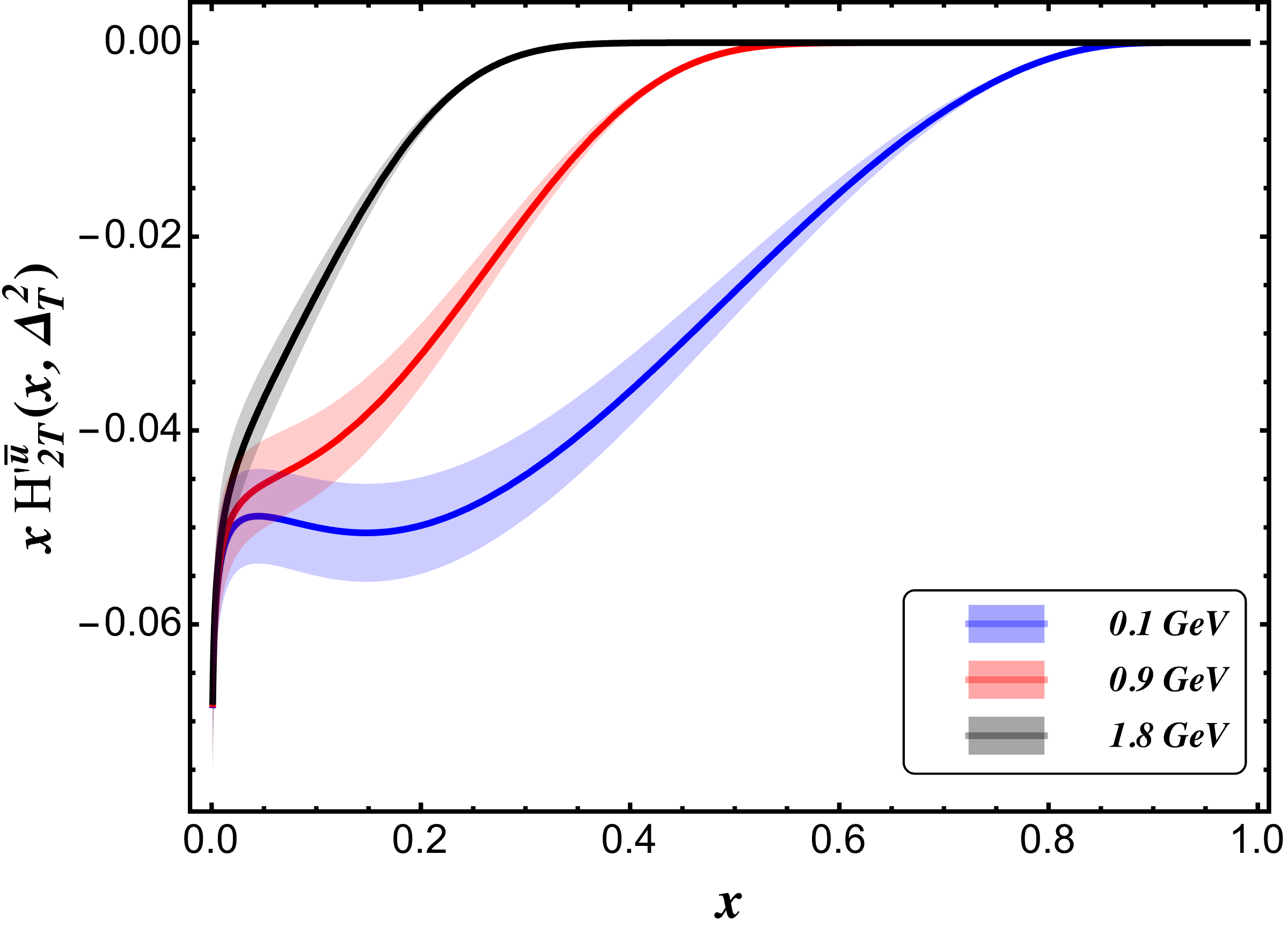} 
        \caption{}
    \end{subfigure}
    \hspace{1cm}
    \begin{subfigure}[b]{0.4\textwidth}
        \centering
        \includegraphics[width=\textwidth]{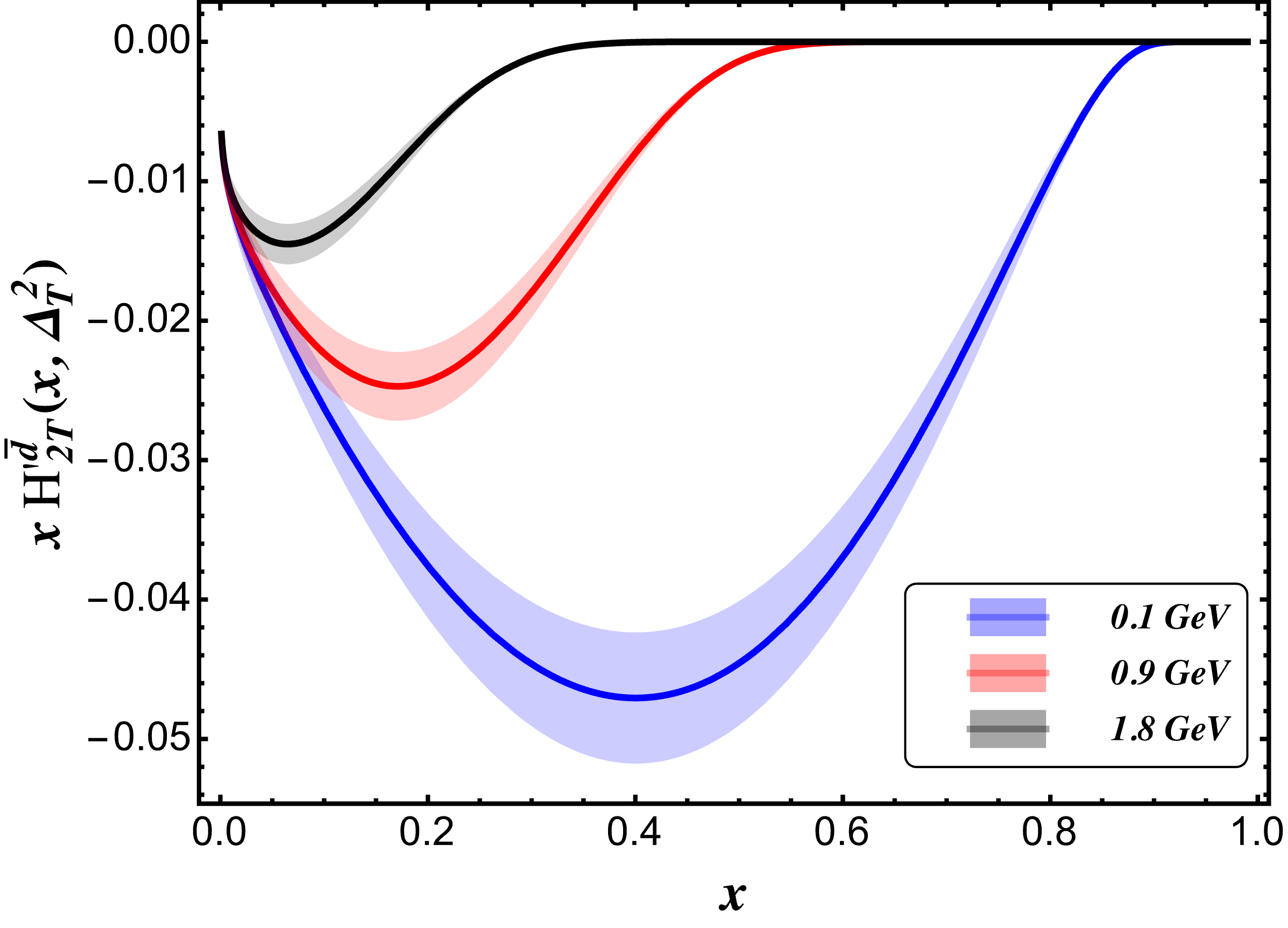} 
        \caption{}
    \end{subfigure}

    \vspace{1em}
    \begin{subfigure}[b]{0.4\textwidth}
        \centering
        \includegraphics[width=\textwidth]{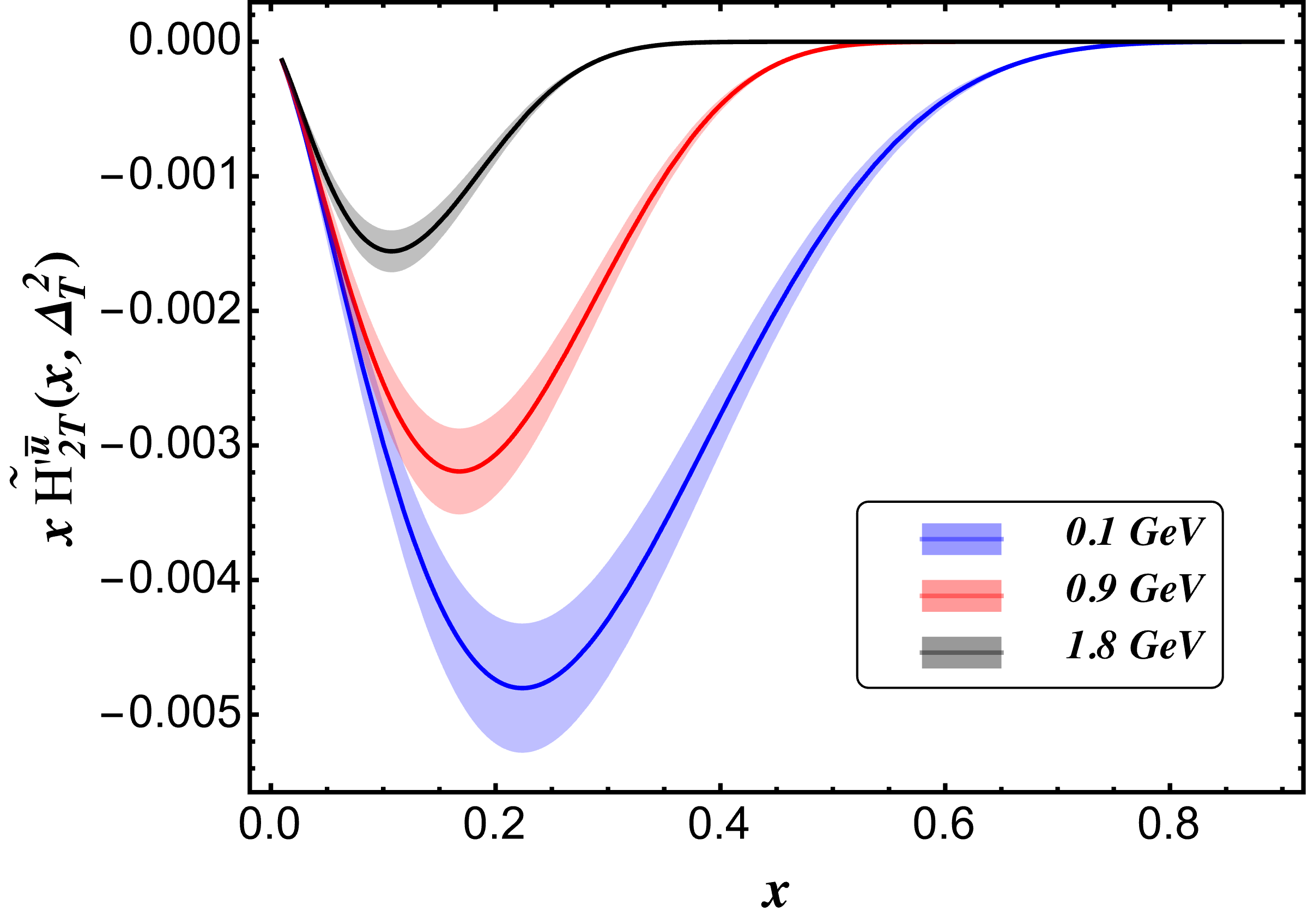} 
        \caption{}
    \end{subfigure}
    \hspace{1cm}
    \begin{subfigure}[b]{0.4\textwidth}
        \centering
        \includegraphics[width=\textwidth]{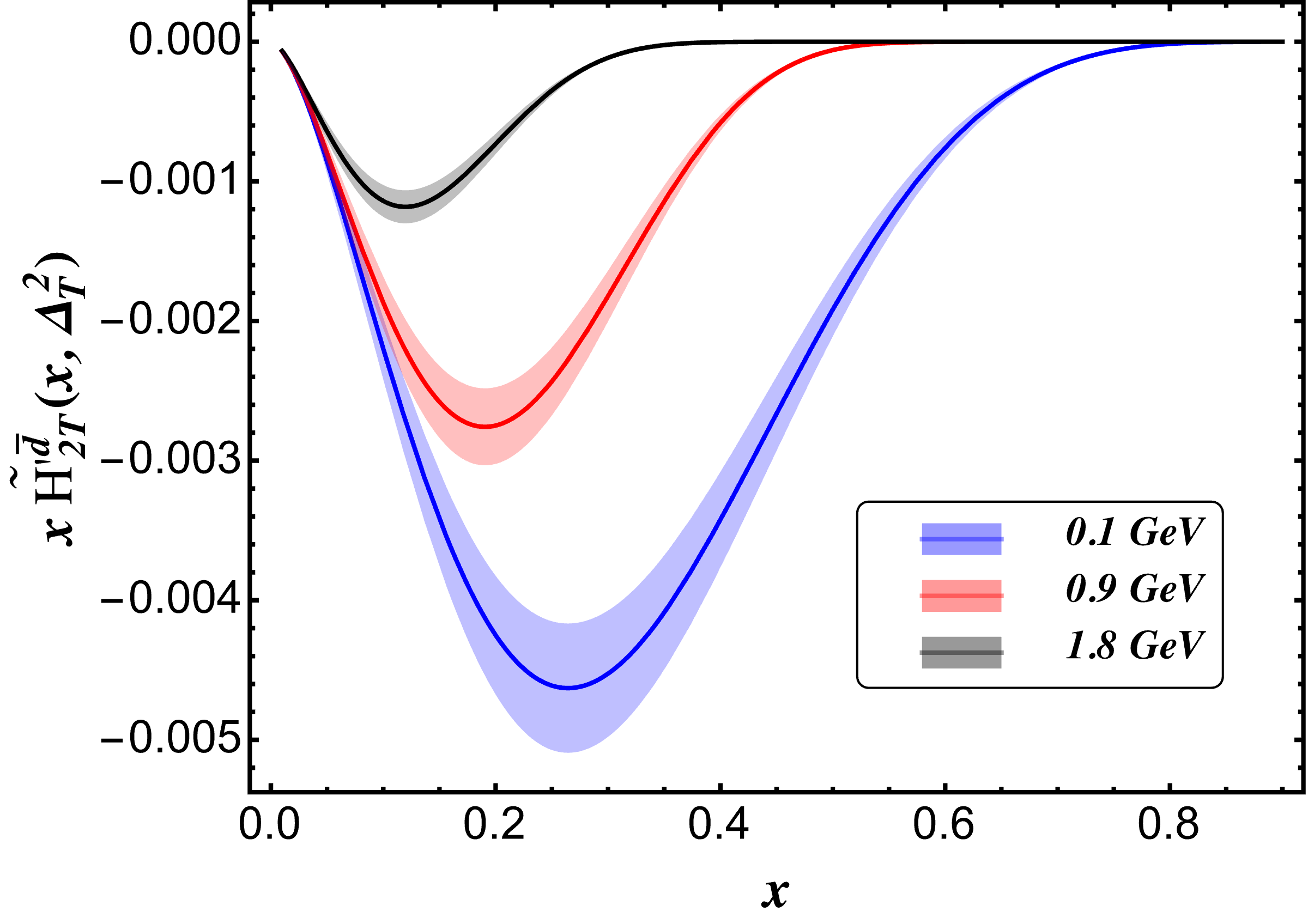} 
        \caption{}
    \end{subfigure}
    \caption{The twist-3 chiral-even generalized parton distributions (GPDs) \(x \tilde{H}_{2T}^{\prime}\), \(x E_{2T}^{\prime}\), and \(x H_{2T}^{\prime}\), corresponding to the Dirac structure \(\gamma^i \gamma_5\), are shown as functions of \(x\) for fixed values of transverse momentum transfer: \(\boldsymbol{\Delta}_T = 0.1 \ \text{GeV}\) (blue), \(0.9 \ \text{GeV}\) (red), and \(1.8 \ \text{GeV}\) (black).}
    \label{figure6}
\end{figure}

A similar analysis is performed for the chiral-even GPDs corresponding to the 
\(\gamma^i\gamma_5\) Dirac structure, specifically \(xE_{2T}^{\prime}\), 
\(xH_{2T}^{\prime}\), and \(x\tilde{H}_{2T}^{\prime}\), as shown in 
Fig.~\ref{figure6}. Subfigures~\ref{figure6}(a) and \ref{figure6}(b) display the 
distributions of \(xE_{2T}^{\prime}\) for the \(\bar{u}\) and \(\bar{d}\) quarks, 
respectively. For the \(\bar{u}\) quark, the distribution exhibits a falloff at small 
\(x\) and large transverse momentum transfer \(\boldsymbol{\Delta}_T\). In contrast, 
the \(\bar{d}\) quark distribution initially rises as \(x\) increases, reaches a 
peak around intermediate \(x\), and then decreases for larger \(x\) values. With 
increasing \(\boldsymbol{\Delta}_T\), the peak becomes less pronounced and eventually 
vanishes at high \(\boldsymbol{\Delta}_T\).  Subfigures~\ref{figure6}(c) and \ref{figure6}(d) show the distribution of 
\(xH_{2T}^{\prime}\), where the \(\bar{u}\) quark contribution dominates over 
\(\bar{d}\) at small \(x\). Moreover, \(xH_{2T}^{\prime}\) increases with increasing 
\(\boldsymbol{\Delta}_T\) and approaches zero at large \(\boldsymbol{\Delta}_T\).  The distributions of \(x\tilde{H}_{2T}^{\prime}\) are shown in 
subfigures~\ref{figure6}(e) and \ref{figure6}(f). Unlike the previous cases, 
\(x\tilde{H}_{2T}^{\prime}\) remains negative throughout the full \(x\)-range and 
for all values of \(\boldsymbol{\Delta}_T\). At small \(x\), both sea quark flavors 
approach zero, with the magnitude of the distribution increasing as 
\(\boldsymbol{\Delta}_T\) decreases. Additionally, the magnitude of the 
\(\bar{u}\) quark contribution exceeds that of the \(\bar{d}\) quark across the 
kinematic region.

\subsection{PDF limit}

Parton distribution functions (PDFs) represent the probability densities for finding a parton carrying a longitudinal momentum fraction \(x\) of the parent hadron. They are formally defined through the quark–quark correlator,

\begin{equation}
    \mathcal{F}(x)=\int \frac{dy^-}{2 \pi} e^{i y^{-} x} \langle P,\Lambda| \Bar{\psi}\bigg(-\frac{y}{2}\bigg)\Gamma \psi \bigg(\frac{y}{2}\bigg)|P,\Lambda \rangle
\end{equation}

In this work, we focus solely on twist-3 chiral-even GPDs. Therefore, we consider the structures \(\Gamma = \gamma^i\) and \(\gamma^i \gamma_5\). In the forward limit and at zero skewness, there is only one chiral-even GPD that has a PDF structure. This is given by:

\begin{equation}
    \int \frac{dy^-}{2 \pi} e^{i y^- x} \langle P,\Lambda| \Bar{\psi}\bigg(-\frac{y}{2}\bigg)\gamma^i \gamma_5 \psi \bigg(\frac{y}{2}\bigg)|P,\Lambda \rangle=2 g_T(x)
\end{equation}

and $g_T$ is connected to the twist-3 GPD $H^{\prime}_{2T}$. The expression that connects the two is given by,
\begin{equation}
    g_T(x)= \lim_{\Delta\to 0} H^{\prime}_{2T}(x,0,-t)
\label{equationpdf}  
\end{equation}

\begin{figure}[htbp]
    \centering
    \begin{subfigure}[b]{0.45\textwidth}
        \centering
        \includegraphics[width=\textwidth]{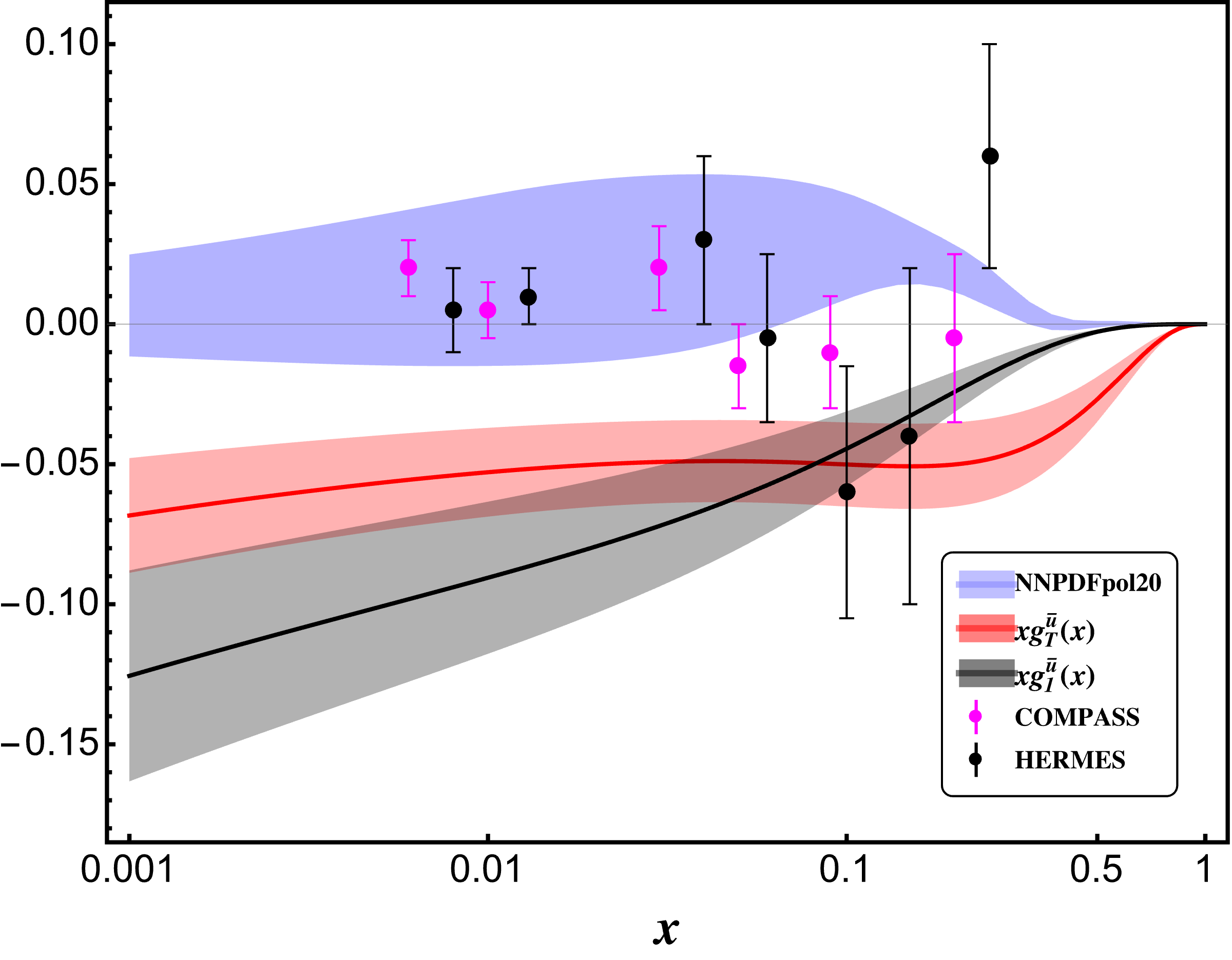} 
        \caption{}
        \label{}
    \end{subfigure}
    \hspace{1cm}
    \begin{subfigure}[b]{0.45\textwidth}
        \centering
        \includegraphics[width=\textwidth]{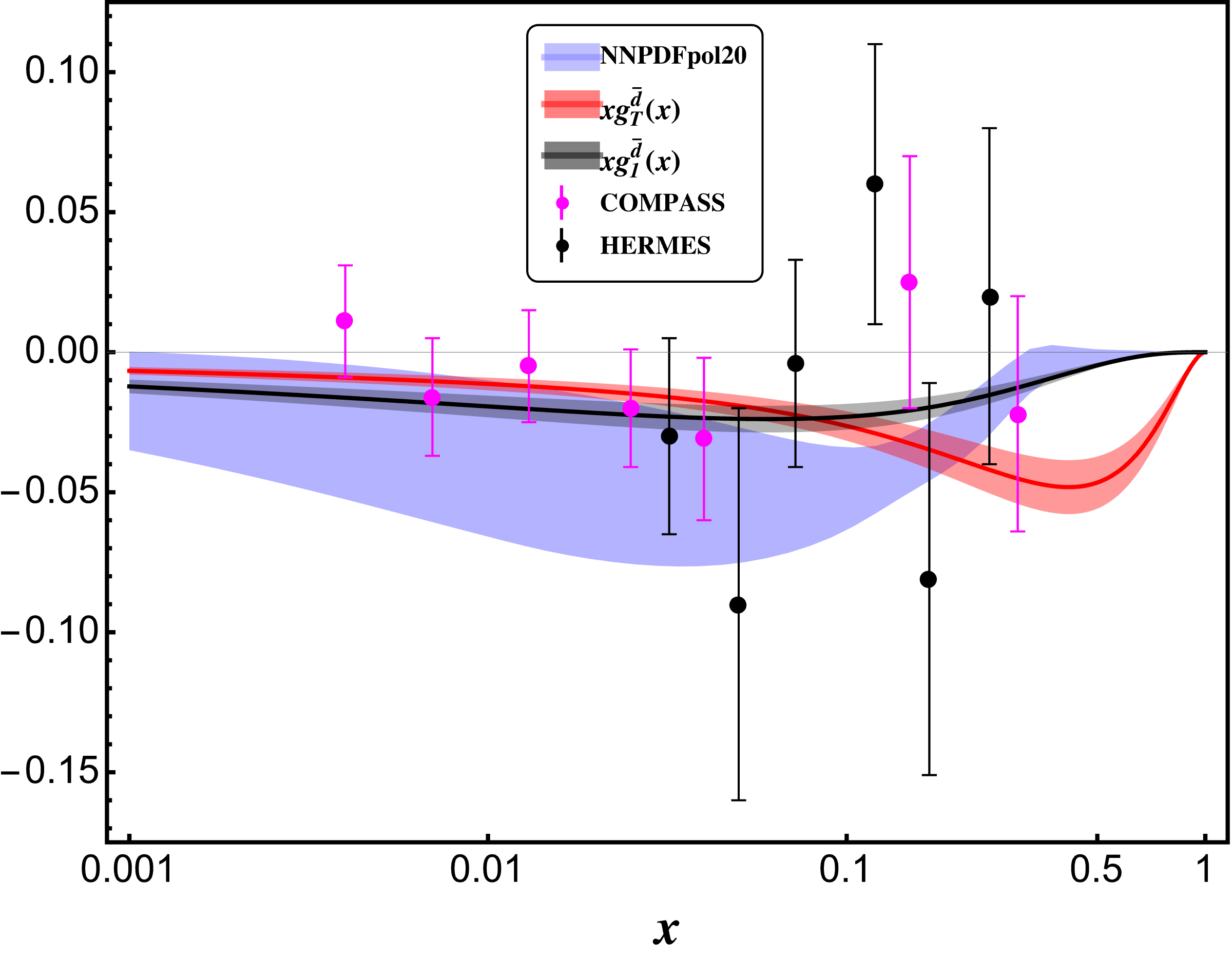} 
        \caption{}
        \label{}
    \end{subfigure}
    \caption{The twist-3 parton distribution function \(x g_T(x)\) and \(xg_1(x)\) are plotted as a function of \(x\) over the range \(0.001 < x < 1\). Panel \(a\) represents the distribution for the \(\bar{u}\) quark, while panel \(b\) corresponds to the \(\bar{d}\) quark. The parton distribution functions are also compare with results from NNPDF~\cite{cite-key}, COMPASS~\cite{alekseev2010quark}, and HERMES~\cite{airapetian2004flavor}.}
    \label{figure7}
\end{figure}

Equations~(\ref{H2prim2T}) and~(\ref{equationpdf}) illustrate that the twist-3 parton distribution function \(g_T(x)\) characterizes the transverse spin distribution of quarks. However, the experimental determination of \(g_T(x)\) remains highly challenging due to its twist-3 nature and the complexity of accessing quark-gluon correlations. In this study, we have computed the numerical values of \(g_T(x)\) in the forward limit \(\boldsymbol{\Delta}_T \to 0\), and the resulting dependence of \(g_T(x)\) on the momentum fraction \(x\) is presented in figure~\ref{figure7} in the kinematic range \(x\in[0.001,1]\).

A central outcome of our analysis is the observation that the distribution 
\(x\,g_T^{\bar{u}}(x)\) exhibits a divergent behavior in the small-\(x\) region, 
while \(x\,g_T^{\bar{d}}(x)\) converges to zero. Notably, both distributions 
remain negative over the considered kinematic range.  The extracted distributions are subsequently compared with available 
results from NNPDF~\cite{cite-key}, COMPASS~\cite{alekseev2010quark}, and 
HERMES~\cite{airapetian2004flavor}. While the \(\bar{d}\) distribution shows 
reasonable agreement with the experimental data, the \(\bar{u}\) distribution 
exhibits a noticeable deviation from the measurements. It is worth emphasizing 
that, although the twist-3 distribution function \(g_T(x)\) lacks a direct 
probabilistic interpretation in terms of parton densities, it nonetheless provides 
essential information on quark--gluon correlations and higher-twist 
effects~\cite{jaffe1992chiral}.

The Burkhardt–Cottingham(BG) sum rule~\cite{burkhardt1970sum} establishes a relation between the twist-3 PDF \(g_T(x)\) and the twist-2 helicity PDF \(g_1(x)\). The sum rule is given by
\begin{equation}
    \int g_2(x) \, dx = 0; \quad \text{with} \quad g_2(x) = g_T(x) - g_1(x).
\end{equation}
Satisfaction of this sum rule implies that the contributions of quark spin to the proton spin remain unchanged under different polarization conditions.

We have numerically evaluated the Burkhardt–Cottingham (BC) sum rule within our model framework. Our results show a violation of the sum rule, which can be attributed to the Fock space truncation employed in the model.
Specifically, the absence of contributions from the valence quark sector and the gluon sector leads to this discrepancy, as the sum rule applies to the full partonic content of the nucleon.

\subsection{Orbital Angular Momentum of Sea Quarks}

In this section, we examine the contribution of sea quarks to the orbital angular momentum (OAM) of the proton. The OAM of quarks is a fundamental component in the decomposition of the nucleon spin and provides valuable insight into the internal transverse dynamics and spin-orbit correlations of partons within the proton.

The OAM associated with a given antiquark flavor \(\bar{\nu}\) can be computed using the following expression~\cite{guo2021novel}:
\begin{align}
    L^{\bar{\nu}}_z(-t=0) = -\frac{1}{2} \int x \, \tilde{E}_{2T}^{\bar{\nu}}(x,0,-t=0) \, dx,
\end{align}
where \(\tilde{E}_{2T}^{\bar{\nu}}(x,0,-t)\) denotes a twist-three generalized parton distribution (GPD) that encodes information about the transverse spin structure of partons. The integration is performed over the longitudinal momentum fraction \(x\), and the squared momentum transfer is defined as \(t = -\boldsymbol{\Delta}_T^2\).

\begin{figure}[htbp]
    \centering
    \begin{subfigure}[b]{0.45\textwidth}
        \centering
        \includegraphics[width=\textwidth]{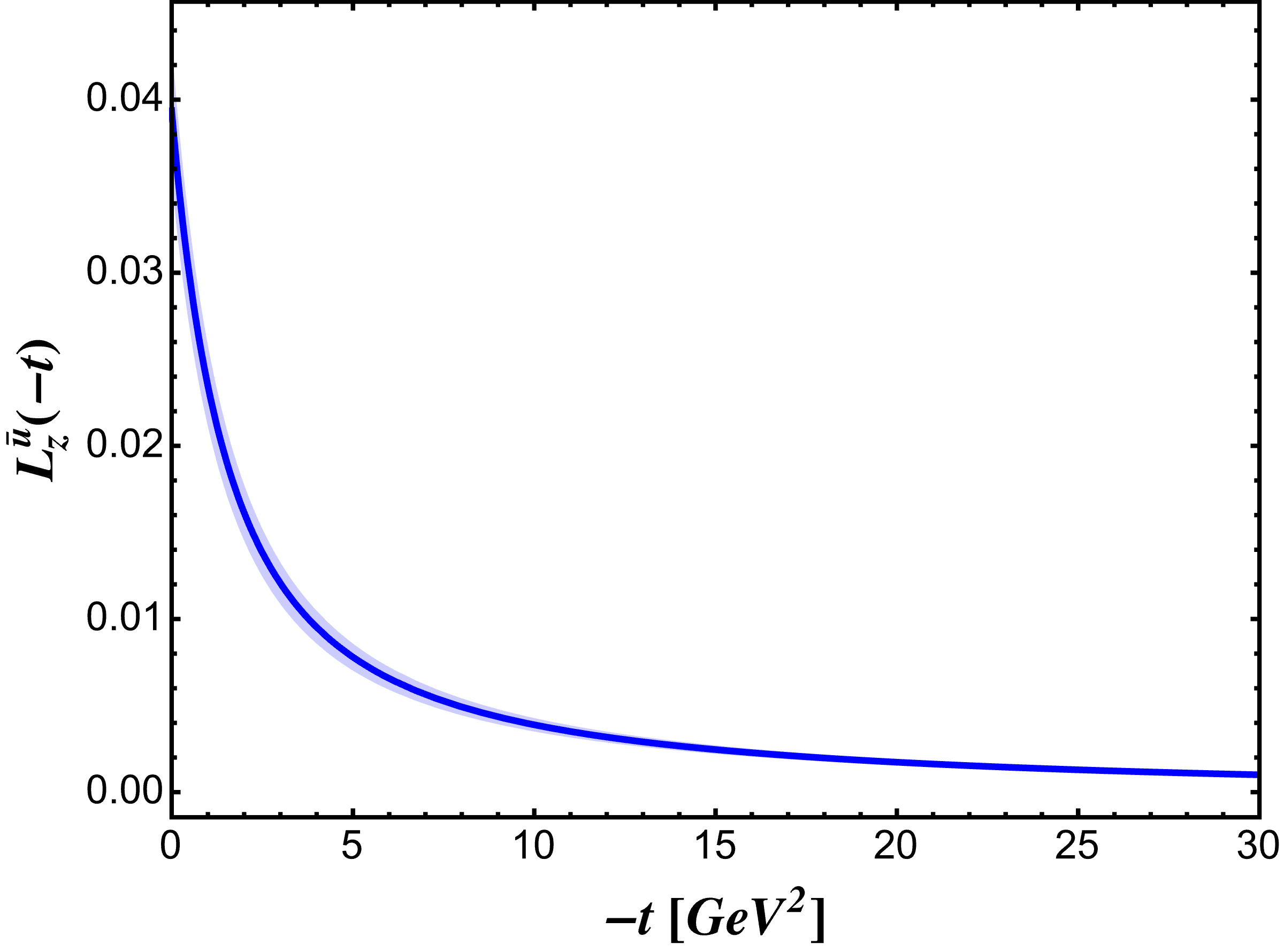} 
        \caption{}
        \label{}
    \end{subfigure}
    \hspace{1cm}
    \begin{subfigure}[b]{0.45\textwidth}
        \centering
        \includegraphics[width=\textwidth]{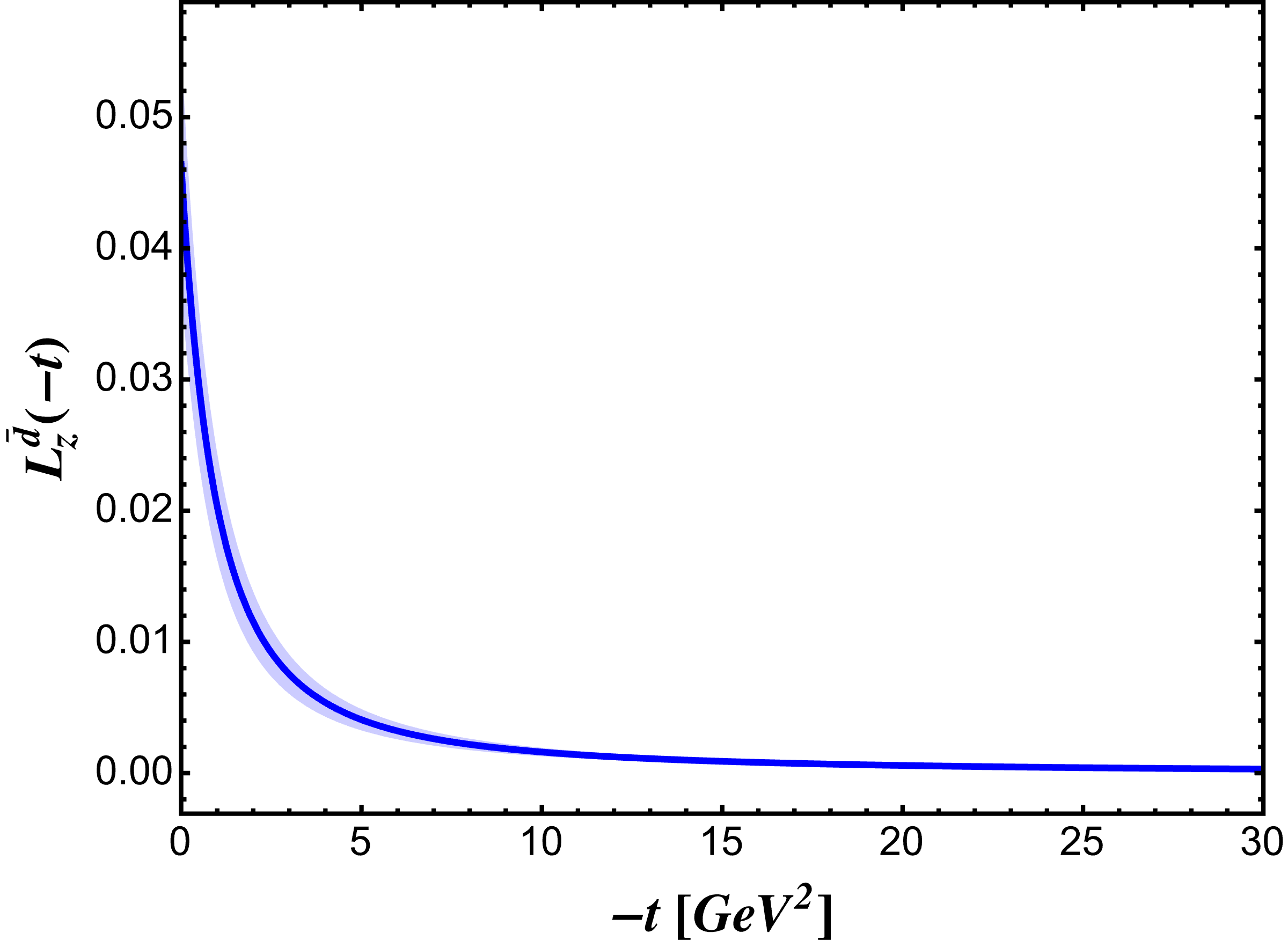} 
        \caption{}
        \label{}
    \end{subfigure}
    \caption{Orbital angular momentum \(L_z^{\bar{\nu}}(-t)\) as a function of transverse momentum transfer square \(\boldsymbol{\Delta}_T^2=-t \ [\text{GeV}^2]\). \((a)\) curve represents the \(\bar{u}\) quark contribution, while \((b)\) corresponds to the \(\bar{d}\) quark.}
    \label{figure8}
\end{figure}

To investigate the dependence of OAM on the transverse momentum transfer, we first numerically integrate over the momentum fraction \(x\), and subsequently analyze the behavior of \(L_z^{\bar{\nu}}(-t)\) as a function of \(\boldsymbol{\Delta}_T\) within the range \(\boldsymbol{\Delta}_T^2 \in [0, 30]\, {\text{GeV}}^2\). The corresponding results are depicted in Figure~\ref{figure8}, where the figure~\ref{figure8}\((a)\) represents the distribution for the \(\bar{u}\) quark, and the figure~\ref{figure8}\((b)\) 
corresponds to the \(\bar{d}\) quark. The plot reveals a convergence of the function with increasing values of \(-t\), indicating the suppression of the OAM contribution at larger transverse momentum transfers.

\begin{table}[hbt]
\centering
\caption{Comparison of the kinetic orbital angular momentum (OAM) calculated in different models at leading twist and twist-3.}
\label{OAM}
\setlength{\tabcolsep}{10pt}
\begin{tabular}{lcc}
\toprule
\textbf{Model} & \textbf{\(L_z^{\bar{u}}\)} & \textbf{\(L_z^{\bar{d}}\)} \\
\midrule
Model-I~\cite{Choudhary:2023unw}   &  \(0.040\pm 0.003\) & \(0.048 \pm 0.002\)  \\
Model-II~\cite{Choudhary:2023unw}   & \(0.048 \pm 0.003\) & \(0.059 \pm 0.002\)  \\
Light-cone model~\cite{Luan:2023lmt}  &  \(0.025\) & \(0.046\) \\
Light-cone model(Twist-3)~\cite{34lm-pr91}  & \(0.024\) & \(0.046\)  \\
Our Model(Twist-3)  & \(0.040\pm 0.001\) & \(0.047 \pm 0.0003\) \\
\bottomrule
\end{tabular}
\end{table}

In Table~\ref{OAM}, we present a comparison of kinetic orbital angular momentum (OAM) 
values calculated in different models for the light sea quarks.  
Our result for \(L^{\bar{d}}_z\) shows good agreement with both Model-I~\cite{Choudhary:2023unw} 
and the Light-Cone Model~\cite{Luan:2023lmt}, as well as the Light-Cone Model (Twist-3)~\cite{34lm-pr91}.  
For \(L^{\bar{u}}_z\), our value aligns more closely with Model-I~\cite{Choudhary:2023unw} and 
Model-II~\cite{Choudhary:2023unw}, while a significant difference is observed compared to the result 
obtained in the Light-Cone Model (Twist-3)~\cite{34lm-pr91}.

\section{CONCLUSION}
\label{Conclusion}

In this paper, we investigate the twist-3 chiral-even generalized parton distributions (GPDs) of light sea quarks, specifically \(\bar{u}\) and \(\bar{d}\), within the light-front spectator model framework. Using the parameterizations from \cite{stepanyan2001observation, rajan2018lorentz}, the GPDs are expressed through light-front wave functions. Our approach closely follows the basis light-front quantization (BLFQ) method as detailed in~\cite{zhang2024twist}.  
Since the BLFQ method is employed only for the calculation of valence quark twist-3 GPDs, a direct comparison with our results for the sea quark sector is not feasible. We further analyze the sea quark asymmetries \(\bar{d}(x) - \bar{u}(x)\) and \(\bar{d}(x)/\bar{u}(x)\) to test the consistency of the model. A thorough study of the two-dimensional (2D) and three-dimensional (3D) distributions of the GPDs is presented, illustrating their dependence on the transverse momentum transfer \(\boldsymbol{\Delta}_T\) and the longitudinal momentum fraction \(x\). We explore the twist-3 parton distribution function (PDF) \(g_T(x)\) for sea quarks alongside the twist-2 PDF \(g_1(x)\), and analyze the Burkhardt--Cottingham sum rule. Our results indicate a violation of this sum rule at small \(x\) values. To validate the model, we compute the contribution of sea quarks to the proton's orbital angular momentum (OAM) and observe good agreement with existing results from both twist-3 and twist-2 model calculations.

Although higher-twist GPDs are suppressed by \(1/Q\) and their direct experimental detection at high-energy facilities such as the Electron-Ion Collider (EIC) and the Electron-Ion Collider in China (EicC) appears highly challenging, our work provides theoretical predictions that contribute to a more complete understanding of nonperturbative parton correlations. In future studies, we plan to compute twist-3 GPDs for sea quarks at non-zero skewness (\(\xi \neq 0\)), including the effects of higher Fock states. This is particularly relevant for the twist-3 cross section of deeply virtual Compton scattering (DVCS)~\cite{guo2022twist}, which will be investigated in upcoming experimental programs at the EIC and EicC~\cite{diehl2001generalized}.  
Ultimately, the study of higher-twist GPDs aims to provide qualitative insights into the internal structure of the proton and contribute to resolving fundamental questions such as the proton spin puzzle and the proton mass problem.

\appendix
\section{}
\label{appendix}

The parametrization of the quark-quark correlator is given in equations \ref{eqnpara} and \ref{eqnpara2}. According to the initial and final helicity of the particle, we can reduce these equations. For $j=1$, we have

\begin{align}
    F^{\gamma^1}_{[++]} &= \frac{\Delta^1}{2 P^+}(E_{2T}+2 \tilde{H}_{2T}) 
    + \frac{i \Delta^2}{2P^+} (\tilde{E}_{2T}-\xi E_{2T})\label{eqn5}  \\
    F^{\gamma^1}_{[--]} &= \frac{\Delta^1}{2 P^+}(E_{2T}+2 \tilde{H}_{2T}) 
    - \frac{i \Delta^2}{2P^+} (\tilde{E}_{2T}-\xi E_{2T})
\end{align}    
\begin{align}    
    F^{\gamma^1 \gamma_5}_{[++]} &= \frac{i \Delta^2}{2 P^+}(E^{\prime}_{2T}+2 \tilde{H}{}_{2T}^{\prime}) 
    - \frac{\Delta^1}{2 P^+}(\tilde{E}^{\prime}_{2T}-\xi E^{\prime}_{2T}) \\ 
    F^{\gamma^1 \gamma_5}_{[--]} &= \frac{i \Delta^2}{2 P^+}(E^{\prime}_{2T}+2 \tilde{H}{}_{2T}^{\prime}) 
    + \frac{\Delta^1}{2 P^+}(\tilde{E}^{\prime}_{2T}-\xi E^{\prime}_{2T}) 
\end{align}
\begin{align}
    F^{\gamma^1}_{[+-]} &= -\frac{M}{P^+}H_{2T} 
    - \frac{(\Delta^1+ i \Delta^2)\Delta^1}{2MP^+}\tilde{H}_{2T}\hspace{1.5cm}\\
    F^{\gamma^1}_{[-+]} &= \frac{M}{P^+}H_{2T} 
    - \frac{(-\Delta^1+ i \Delta^2)\Delta^1}{2MP^+}\tilde{H}_{2T}\hspace{1.5cm}
\end{align}
\begin{align}
    F^{\gamma^1 \gamma_5}_{[+-]} &= \frac{M}{P^+}H{}^{\prime}_{2T} 
    - \frac{i(\Delta^1+i \Delta^2)\Delta^2}{2MP^+}\tilde{H}{}^{\prime}_{2T}\hspace{1cm}  \\
     F^{\gamma^1 \gamma_5}_{[-+]} &= -\frac{M}{P^+}H{}^{\prime}_{2T} 
    - \frac{i(-\Delta^1+i \Delta^2)\Delta^2}{2MP^+}\tilde{H}{}^{\prime}_{2T}\hspace{1cm}
\end{align}

Similarly for $j=2$ we have the following,

\begin{align}
    F^{\gamma^2}_{[++]} &= \frac{\Delta^2}{2 P^+}(E_{2T}+2 \tilde{H}_{2T}) 
    - \frac{i \Delta^1}{2P^+} (\tilde{E}_{2T}-\xi E_{2T}) \\
    F^{\gamma^2}_{[--]} &= \frac{\Delta^2}{2 P^+}(E_{2T}+2 \tilde{H}_{2T}) + \frac{i \Delta^1}{2P^+}(\tilde{E}_{2T}-\xi E_{2T})
\end{align}
\begin{align}
    F^{\gamma^2 \gamma_5}_{[++]} &= -\frac{i \Delta^1}{2 P^+}(E^{\prime}_{2T}+2 \tilde{H}{}_{2T}^{\prime}) 
    - \frac{\Delta^2}{2 P^+}(\tilde{E}^{\prime}_{2T}-\xi E^{\prime}_{2T})\\
    F^{\gamma^2 \gamma_5}_{[--]} &=- \frac{i \Delta^1}{2 P^+} (E^{\prime}_{2T}+2 \tilde{H}{}_{2T}^{\prime}) 
    + \frac{\Delta^2}{2 P^+}(\tilde{E}^{\prime}_{2T}-\xi E^{\prime}_{2T})\\
    F^{\gamma^2}_{[+-]} &= -\frac{i M}{P^+}H_{2T} 
    - \frac{(\Delta^1+ i \Delta^2)\Delta^2}{2MP^+}\tilde{H}_{2T}\\
    F^{\gamma^2}_{[-+]} &= -\frac{i M}{P^+}H_{2T} 
    - \frac{(-\Delta^1+ i \Delta^2)\Delta^2}{2MP^+}\tilde{H}_{2T} \\
    F^{\gamma^2 \gamma_5}_{[+-]} &= \frac{iM}{P^+}H{}^{\prime}_{2T} 
    - \frac{i(\Delta^1+i \Delta^2)\Delta^1}{2MP^+}\tilde{H}{}^{\prime}_{2T}\\ 
    F^{\gamma^2 \gamma_5}_{[-+]} &= -\frac{iM}{P^+}H{}^{\prime}_{2T} 
    - \frac{i(-\Delta^1+i \Delta^2)\Delta^1}{2MP^+}\tilde{H}{}^{\prime}_{2T}
\label{eqn12}    
\end{align}

\section{}

The overlap representation of the chiral even twist-3 
GPDs of sea quarks in terms of the LCWFs can be
expressed as

\begin{align}
i\frac{\boldsymbol{\Delta}_T^2}{2}\tilde{E}_{2T}(x, 0, t) &= -\sum_{\lambda_{q_f} \lambda_{q_i}}\int \frac{d^2\boldsymbol{k}_T}{16\pi^3}  
\bigg\{ \Delta_2 \bigg[ \frac{2k_1 + i\Delta_2}{4x} (\psi_{\lambda_A\lambda_{q_f}}^{+*} \psi_{\lambda_A\lambda_{q_i}}^+ \notag\\
&- \psi_{\lambda_A\lambda_{q_f}}^{-*} \psi_{\lambda_A\lambda_{q_i}}^-) + \frac{2k_1 - i\Delta_2}{4x} (\psi_{\lambda_A\lambda_{q_f}}^{+*} \psi_{\lambda_A\lambda_{q_i}}^+ + \psi_{\lambda_A\lambda_{q_f}}^{-*} \psi_{\lambda_A\lambda_{q_i}}^-) \bigg]  \notag\\
&- \Delta_1 \bigg[ \frac{2k_2 - i\Delta_1}{4x} (\psi_{\lambda_A\lambda_{q_f}}^{+*} \psi_{\lambda_A\lambda_{q_i}}^+ - \psi_{\lambda_A\lambda_{q_f}}^{-*} \psi_{\lambda_A\lambda_{q_i}}^-) \notag\\
&+ \frac{2k_2 + i\Delta_1}{4x} (\psi_{\lambda_A\lambda_{q_f}}^{+*} \psi_{\lambda_A\lambda_{q_i}}^+ + \psi_{\lambda_A\lambda_{q_f}}^{-*} \psi_{\lambda_A\lambda_{q_i}}^-) \bigg] \bigg\},
\end{align}

\begin{align}
&i\frac{\boldsymbol{\Delta}_T^2}{2}({E}'_{2T} + 2\tilde{H}'_{2T})\notag\\
&=  \int \frac{d^2\boldsymbol{k}_T}{16\pi^3}  \sum_{\lambda_{q_f} \lambda_{q_i}} \Biggl\{ \Delta_2 \Biggl[ \frac{m}{2x} (\psi_{\lambda_A\lambda_{q_f}}^{+*} \psi_{\lambda_A\lambda_{q_i}}^+ + \psi_{\lambda_A\lambda_{q_f}}^{-*} \psi_{\lambda_A\lambda_{q_i}}^- + \psi_{\lambda_A\lambda_{q_f}}^{+*} \psi_{\lambda_A\lambda_{q_i}}^+  \notag\\
&- \psi_{\lambda_A\lambda_{q_f}}^{-*} \psi_{\lambda_A\lambda_{q_i}}^-)+ \frac{-2k_1 + i\Delta_2}{4x} (\psi_{\lambda_A\lambda_{q_f}}^{+*} \psi_{\lambda_A\lambda_{q_i}}^+ + \psi_{\lambda_A\lambda_{q_f}}^{-*} \psi_{\lambda_A\lambda_{q_i}}^-) + \frac{2k_1 + i\Delta_2}{4x}\notag \\
&(\psi_{\lambda_A\lambda_{q_f}}^{+*} \psi_{\lambda_A\lambda_{q_i}}^- + \psi_{\lambda_A\lambda_{q_f}}^{-*} \psi_{\lambda_A\lambda_{q_i}}^-) \Biggr] - \Delta_1 \Biggl[ \frac{im}{2x} (-\psi_{\lambda_A\lambda_{q_f}}^{+*} \psi_{\lambda_A\lambda_{q_i}}^+ - \psi_{\lambda_A\lambda_{q_f}}^{-*} \psi_{\lambda_A\lambda_{q_i}}^- \notag\\
&+ \psi_{\lambda_A\lambda_{q_f}}^{+*} \psi_{\lambda_A\lambda_{q_i}}^+ + \psi_{\lambda_A\lambda_{q_f}}^{-*} \psi_{\lambda_A\lambda_{q_i}}^-)+ \frac{-2k_2 - i\Delta_1}{4x} (\psi_{\lambda_A\lambda_{q_f}}^{+*} \psi_{\lambda_A\lambda_{q_i}}^+ + \psi_{\lambda_A\lambda_{q_f}}^{-*} \psi_{\lambda_A\lambda_{q_i}}^-)\notag \\
&+ \frac{2k_2 - i\Delta_1}{4x} (\psi_{\lambda_A\lambda_{q_f}}^{+*} \psi_{\lambda_A\lambda_{q_i}}^- + \bar{\psi}_{\lambda_A\lambda_{q_f}}^{-*} \psi_{\lambda_A\lambda_{q_i}}^-) \Biggr] \Biggr\},
\end{align}

\begin{align}
&H'_{2T} + \frac{\boldsymbol{\Delta}_T^2}{4M^2} \tilde{H}'_{2T} \notag\\
&=  \int \frac{d^2\boldsymbol{k}_T}{16\pi^3} \sum_{\lambda_{q_f} \lambda_{q_i}} \Biggl\{ \Biggl[ \frac{m}{4Mx} (\psi_{\lambda_A\lambda_{q_f}}^{+*} \psi_{\lambda_A\lambda_{q_i}}^{-} + \psi_{\lambda_A\lambda_{q_f}}^{-*} \psi_{\lambda_A\lambda_{q_i}}^{+} + \psi_{\lambda_A\lambda_{q_f}}^{-*} \psi_{\lambda_A\lambda_{q_i}}^{-} +  \notag\\
&\psi_{\lambda_A\lambda_{q_f}}^{+*} \psi_{\lambda_A\lambda_{q_i}}^{-})+ \frac{-2k_1 + i\Delta_2}{8Mx} (\psi_{\lambda_A\lambda_{q_f}}^{+*} \psi_{\lambda_A\lambda_{q_i}}^- + \psi_{\lambda_A\lambda_{q_f}}^{-*} \psi_{\lambda_A\lambda_{q_i}}^+) + \frac{2k_1 + i\Delta_2}{8Mx} \notag\\
&(\psi_{\lambda_A\lambda_{q_f}}^{+*} \psi_{\lambda_A\lambda_{q_i}}^- + \psi_{\lambda_A\lambda_{q_f}}^{-*} \psi_{\lambda_A\lambda_{q_i}}^-) \Biggr] + i \Biggl[ \frac{im}{4Mx} (-\psi_{\lambda_A\lambda_{q_f}}^{+*} \psi_{\lambda_A\lambda_{q_i}}^- + \psi_{\lambda_A\lambda_{q_f}}^{-*} \psi_{\lambda_A\lambda_{q_i}}^+ \notag\\
&+ \psi_{\lambda_A\lambda_{q_f}}^{+*} \psi_{\lambda_A\lambda_{q_i}}^- - \psi_{\lambda_A\lambda_{q_f}}^{+*} \psi_{\lambda_A\lambda_{q_i}}^-) + \frac{-2k_2 - i\Delta_1}{8Mx} (\psi_{\lambda_A\lambda_{q_f}}^{+*} \psi_{\lambda_A\lambda_{q_i}}^+ - \psi_{\lambda_A\lambda_{q_f}}^{-*} \psi_{\lambda_A\lambda_{q_i}}^+) \notag\\
&+ \frac{2k_2 - i\Delta_1}{8Mx} (\psi_{\lambda_A\lambda_{q_f}}^{+*} \psi_{\lambda_A\lambda_{q_i}}^- - \psi_{\lambda_A\lambda_{q_f}}^{-*} \psi_{\lambda_A\lambda_{q_i}}^-) \Biggr] \Biggr\},
\end{align}

\begin{align}
&i\frac{\Delta_1\Delta_2}{M}\tilde{H}'_{2T}(x, 0, t) \notag\\
= & \int \frac{d^2\boldsymbol{k}_T}{16\pi^3} \sum_{\lambda_{q_f} \lambda_{q_i}}  \Biggl\{ \Biggl[ \frac{m}{2x} (\psi_{\lambda_A\lambda_{q_f}}^{+*} \psi_{\lambda_A\lambda_{q_i}}^- - \psi_{\lambda_A\lambda_{q_f}}^{-*} \psi_{\lambda_A\lambda_{q_i}}^+ + \psi_{\lambda_A\lambda_{q_f}}^{+*} \psi_{\lambda_A\lambda_{q_i}}^- - \notag \\
&\psi_{\lambda_A\lambda_{q_f}}^{-*} \psi_{\lambda_A\lambda_{q_i}}^+)+ \frac{-2k_1 + i\Delta_2}{4x} (\psi_{\lambda_A\lambda_{q_f}}^{+*} \psi_{\lambda_A\lambda_{q_i}}^+ - \psi_{\lambda_A\lambda_{q_f}}^{-*} \psi_{\lambda_A\lambda_{q_i}}^+) + \frac{2k_1 + i\Delta_2}{4x}  \notag\\
&(\psi_{\lambda_A\lambda_{q_f}}^{+*} \psi_{\lambda_A\lambda_{q_i}}^- - \psi_{\lambda_A\lambda_{q_f}}^{-*} \psi_{\lambda_A\lambda_{q_i}}^+) \Biggr] - i \Biggl[ \frac{im}{2x} (-\psi_{\lambda_A\lambda_{q_f}}^{+*} \psi_{\lambda_A\lambda_{q_i}}^- - \psi_{\lambda_A\lambda_{q_f}}^{-*} \psi_{\lambda_A\lambda_{q_i}}^+ \notag \\
& + \psi_{\lambda_A\lambda_{q_f}}^{+*} \psi_{\lambda_A\lambda_{q_i}}^- + \psi_{\lambda_A\lambda_{q_f}}^{-*} \psi_{\lambda_A\lambda_{q_i}}^+) + \frac{-2k_2 - i\Delta_1}{4x} (\psi_{\lambda_A\lambda_{q_f}}^{+*} \psi_{\lambda_A\lambda_{q_i}}^+ \notag \\
&+ \psi_{\lambda_A\lambda_{q_f}}^{-*} \psi_{\lambda_A\lambda_{q_i}}^+)+ \frac{2k_2 - i\Delta_1}{4x} (\psi_{\lambda_A\lambda_{q_f}}^{+*} \psi_{\lambda_A\lambda_{q_i}}^- + \psi_{\lambda_A\lambda_{q_f}}^{-*} \psi_{\lambda_A\lambda_{q_i}}^+) \Biggr] \Biggr\}.
\end{align}

\nocite{*}
\bibliographystyle{JHEP}
\bibliography{ref}

\providecommand{\href}[2]{#2}\begingroup\raggedright\begin{thebibliography}{10}

\bibitem{deur2019spin}
A.~Deur, S.J.~Brodsky and G.F.~De~T{\'e}ramond, \emph{The spin structure of the nucleon}, {\emph{Reports on Progress in Physics} {\bfseries 82} (2019) 076201}.

\bibitem{boffi2007generalized}
S.~Boffi and B.~Pasquini, \emph{Generalized parton distributions and the structure of the nucleon}, {\emph{La Rivista del Nuovo Cimento} {\bfseries 30} (2007) 387}.

\bibitem{kuhn2009spin}
S.~Kuhn, J.-P.~Chen and E.~Leader, \emph{Spin structure of the nucleon—status and recent results}, {\emph{Progress in Particle and Nuclear Physics} {\bfseries 63} (2009) 1}.

\bibitem{filippone2001spin}
B.~Filippone and X.~Ji, \emph{The spin structure of the nucleon},  in \emph{Advances in nuclear physics}, pp.~1--88, Springer (2001).

\bibitem{aidala2013spin}
C.A.~Aidala, S.D.~Bass, D.~Hasch and G.K.~Mallot, \emph{The spin structure of the nucleon}, {\emph{Reviews of Modern Physics} {\bfseries 85} (2013) 655}.

\bibitem{ji1997gauge}
X.~Ji, \emph{Gauge-invariant decomposition of nucleon spin}, {\emph{Physical Review Letters} {\bfseries 78} (1997) 610}.

\bibitem{ji1997deeply}
X.~Ji, \emph{Deeply virtual compton scattering}, {\emph{Physical Review D} {\bfseries 55} (1997) 7114}.

\bibitem{radyushkin1996scaling}
A.~Radyushkin, \emph{Scaling limit of deeply virtual compton scattering}, {\emph{Physics Letters B} {\bfseries 380} (1996) 417}.

\bibitem{belitsky2002theory}
A.V.~Belitsky, D.~Mueller and A.~Kirchner, \emph{Theory of deeply virtual compton scattering on the nucleon}, {\emph{Nuclear Physics B} {\bfseries 629} (2002) 323}.

\bibitem{Hashamipour:2020kip}
H.~Hashamipour, M.~Goharipour and S.S.~Gousheh, \emph{{Determination of generalized parton distributions through a simultaneous analysis of axial form factor and wide-angle Compton scattering data}}, \href{https://doi.org/10.1103/PhysRevD.102.096014}{\emph{Phys. Rev. D} {\bfseries 102} (2020) 096014} [\href{https://arxiv.org/abs/2006.05760}{{\ttfamily 2006.05760}}].

\bibitem{goloskokov2007longitudinal}
S.~Goloskokov and P.~Kroll, \emph{The longitudinal cross section of vector meson electroproduction}, {\emph{The European Physical Journal C} {\bfseries 50} (2007) 829}.

\bibitem{goloskokov2008role}
S.~Goloskokov and P.~Kroll, \emph{The role of the quark and gluon gpds in hard vector-meson electroproduction}, {\emph{The European Physical Journal C} {\bfseries 53} (2008) 367}.

\bibitem{goloskokov2010attempt}
S.V.~Goloskokov and P.~Kroll, \emph{An attempt to understand exclusive $\pi$+ electroproduction}, {\emph{The European Physical Journal C} {\bfseries 65} (2010) 137}.

\bibitem{goloskokov2011transversity}
S.~Goloskokov and P.~Kroll, \emph{Transversity in hard exclusive electroproduction of pseudoscalar mesons}, {\emph{The European Physical Journal A} {\bfseries 47} (2011) 112}.

\bibitem{adloff2002diffractive}
C.~Adloff, V.~Andreev, B.~Andrieu, T.~Anthonis, A.~Astvatsatourov, A.~Babaev et~al., \emph{Diffractive photoproduction of $\psi$ (2s) mesons at hera}, {\emph{Physics Letters B} {\bfseries 541} (2002) 251}.

\bibitem{aaron2009deeply}
F.D.~Aaron, M.A.~Martin, C.~Alexa, K.~Alimujiang, V.~Andreev, B.~Antunovic et~al., \emph{Deeply virtual compton scattering and its beam charge asymmetry in e$\pm$p collisions at hera}, {\emph{Physics Letters B} {\bfseries 681} (2009) 391}.

\bibitem{zeus2009measurement}
Z.~Collaboration et~al., \emph{A measurement of the q2, w and t dependences of deeply virtual compton scattering at hera}, {\emph{Journal of High Energy Physics} {\bfseries 2009} (2009) 108}.

\bibitem{airapetian2012beam}
A.~Airapetian, N.~Akopov, Z.~Akopov, E.~Aschenauer, W.~Augustyniak, R.~Avakian et~al., \emph{Beam-helicity and beam-charge asymmetries associated with deeply virtual compton scattering on the unpolarised proton}, {\emph{Journal of High Energy Physics} {\bfseries 2012} (2012) }.

\bibitem{airapetian2012beam1}
A.~Airapetian, N.~Akopov, Z.~Akopov, E.~Aschenauer, W.~Augustyniak, R.~Avakian et~al., \emph{Beam-helicity asymmetry arising from deeply virtual compton scattering measured with kinematically complete event reconstruction}, {\emph{Journal of High Energy Physics} {\bfseries 2012} (2012) 1}.

\bibitem{d2004feasibility}
N.~d’Hose, E.~Burtin, P.~Guichon and J.~Marroncle, \emph{Feasibility study of deeply virtual compton scattering using compass at cern},  in \emph{Perspectives in Hadronic Physics: 4th International Conference Held at ICTP, Trieste, Italy, 12--16 May 2003}, pp.~47--53, Springer, 2004.

\bibitem{stepanyan2001observation}
S.~Stepanyan, V.~Burkert, L.~Elouadrhiri, G.~Adams, E.~Anciant, M.~Anghinolfi et~al., \emph{Observation of exclusive deeply virtual compton scattering in polarized electron beam asymmetry measurements}, {\emph{Physical Review Letters} {\bfseries 87} (2001) 182002}.

\bibitem{Goharipour:2024atx}
{\scshape MMGPDs} collaboration, \emph{{Impact of JLab data on the determination of GPDs at zero skewness and new insights from transition form factors $N\rightarrow \Delta$}}, \href{https://doi.org/10.1103/PhysRevD.109.074042}{\emph{Phys. Rev. D} {\bfseries 109} (2024) 074042} [\href{https://arxiv.org/abs/2403.19384}{{\ttfamily 2403.19384}}].

\bibitem{ji1997study}
X.~Ji, W.~Melnitchouk and X.~Song, \emph{Study of off-forward parton distributions}, {\emph{Physical Review D} {\bfseries 56} (1997) 5511}.

\bibitem{scopetta2004generalized}
S.~Scopetta and V.~Vento, \emph{Generalized parton distributions and composite constituent quarks}, {\emph{Physical Review D} {\bfseries 69} (2004) 094004}.

\bibitem{choi2002continuity}
H.-M.~Choi, C.-R.~Ji and L.~Kisslinger, \emph{Continuity of generalized parton distributions for the pion virtual compton scattering}, {\emph{Physical Review D} {\bfseries 66} (2002) 053011}.

\bibitem{choi2001skewed}
H.-M.~Choi, C.-R.~Ji and L.~Kisslinger, \emph{Skewed quark distribution of the pion in the light-front quark model}, {\emph{Physical Review D} {\bfseries 64} (2001) 093006}.

\bibitem{mineo2005generalized}
H.~Mineo, S.N.~Yang, C.-Y.~Cheung and W.~Bentz, \emph{Generalized parton distributions of the nucleon in the nambu--jona-lasinio model based on the faddeev approach}, {\emph{Physical Review C—Nuclear Physics} {\bfseries 72} (2005) 025202}.

\bibitem{goeke2008generalized}
K.~Goeke, V.~Guzey and M.~Siddikov, \emph{Generalized parton distributions and deeply virtual compton scattering in color glass condensate model}, {\emph{The European Physical Journal C} {\bfseries 56} (2008) 203}.

\bibitem{goeke2001hard}
K.~Goeke, M.V.~Polyakov and M.~Vanderhaeghen, \emph{Hard exclusive reactions and the structure of hadrons}, {\emph{Progress in Particle and Nuclear Physics} {\bfseries 47} (2001) 401}.

\bibitem{ossmann2005generalized}
J.~Ossmann, M.~Polyakov, P.~Schweitzer, D.~Urbano and K.~Goeke, \emph{Generalized parton distribution function (e u+ e d)(x, $\xi$, t) of the nucleon<? format?> in the chiral quark soliton model}, {\emph{Physical Review D—Particles, Fields, Gravitation, and Cosmology} {\bfseries 71} (2005) 034011}.

\bibitem{tiburzi2002exploring}
B.~Tiburzi and G.~Miller, \emph{Exploring skewed parton distributions with two-body models on the light front. ii. covariant bethe-salpeter approach}, {\emph{Physical Review D} {\bfseries 65} (2002) 074009}.

\bibitem{noguera2004generalized}
S.~Noguera, L.~Theu{\ss}l and V.~Vento, \emph{Generalized parton distributions of the pion in a bethe-salpeter approach}, {\emph{The European Physical Journal A-Hadrons and Nuclei} {\bfseries 20} (2004) 483}.

\bibitem{pasquini2006virtual}
B.~Pasquini and S.~Boffi, \emph{Virtual meson cloud of the nucleon and generalized parton distributions}, {\emph{Physical Review D—Particles, Fields, Gravitation, and Cosmology} {\bfseries 73} (2006) 094001}.

\bibitem{pasquini2007generalized}
B.~Pasquini and S.~Boffi, \emph{Generalized parton distributions in a meson cloud model}, {\emph{Nuclear Physics A} {\bfseries 782} (2007) 86}.

\bibitem{jaffe1990g1}
R.~Jaffe and A.~Manohar, \emph{The g1 problem: Deep inelastic electron scattering and the spin of the proton}, {\emph{Nuclear Physics B} {\bfseries 337} (1990) 509}.

\bibitem{hatta2012twist}
Y.~Hatta and S.~Yoshida, \emph{Twist analysis of the nucleon spin in qcd}, {\emph{Journal of high energy physics} {\bfseries 2012} (2012) 1}.

\bibitem{guo2022twist}
Y.~Guo, X.~Ji, B.~Kriesten and K.~Shiells, \emph{Twist-three cross-sections in deeply virtual compton scattering}, {\emph{Journal of High Energy Physics} {\bfseries 2022} (2022) 1}.

\bibitem{burkardt2013transverse}
M.~Burkardt, \emph{Transverse force on quarks in deep-inelastic scattering}, {\emph{Physical Review D—Particles, Fields, Gravitation, and Cosmology} {\bfseries 88} (2013) 114502}.

\bibitem{aslan2019transverse}
F.P.~Aslan, M.~Burkardt and M.~Schlegel, \emph{Transverse force tomography}, {\emph{Physical Review D} {\bfseries 100} (2019) 096021}.

\bibitem{mukherjee2002off}
A.~Mukherjee and M.~Vanderhaeghen, \emph{Off-forward matrix elements in light-front hamiltonian qcd}, {\emph{Physics Letters B} {\bfseries 542} (2002) 245}.

\bibitem{mukherjee2003helicity}
A.~Mukherjee and M.~Vanderhaeghen, \emph{Helicity-dependent twist-two and twist-three generalized parton distributions in light-front qcd}, {\emph{Physical Review D} {\bfseries 67} (2003) 085020}.

\bibitem{aslan2020singularities}
F.P.~Aslan and M.~Burkardt, \emph{Singularities in twist-3 quark distributions}, {\emph{Physical Review D} {\bfseries 101} (2020) 016010}.

\bibitem{Sharma:2023ibp}
S.~Sharma and H.~Dahiya, \emph{{Exploring twist-4 chiral-even GPDs in the light-front quark-diquark model}}, \href{https://doi.org/10.1016/j.nuclphysb.2024.116522}{\emph{Nucl. Phys. B} {\bfseries 1001} (2024) 116522} [\href{https://arxiv.org/abs/2310.03463}{{\ttfamily 2310.03463}}].

\bibitem{bhattacharya2023chiral}
S.~Bhattacharya, K.~Cichy, M.~Constantinou, J.~Dodson, A.~Metz, A.~Scapellato et~al., \emph{Chiral-even axial twist-3 gpds of the proton from lattice qcd}, {\emph{Physical Review D} {\bfseries 108} (2023) 054501}.

\bibitem{zhang2024twist}
Z.~Zhang, Z.~Hu, S.~Xu, C.~Mondal, X.~Zhao, J.P.~Vary et~al., \emph{Twist-3 generalized parton distribution for the proton from basis light-front quantization}, {\emph{Physical Review D} {\bfseries 109} (2024) 034031}.

\bibitem{burkhardt1970sum}
H.~Burkhardt and W.~Cottingham, \emph{Sum rules for forward virtual compton scattering}, {\emph{Annals of Physics} {\bfseries 56} (1970) 453}.

\bibitem{Cloet2014}
I.C.~Cloët and C.D.~Roberts, \emph{Explanation and prediction of observables using continuum strong qcd}, \href{https://doi.org/10.1016/j.ppnp.2014.02.001}{\emph{Prog. Part. Nucl. Phys.} {\bfseries 77} (2014) 1}.

\bibitem{Ellis_2009}
J.~Ellis, D.S.~Hwang and A.~Kotzinian, \emph{Sivers asymmetries for inclusive pion and kaon production in deep-inelastic scattering}, \href{https://doi.org/10.1103/physrevd.80.074033}{\emph{Physical Review D} {\bfseries 80} (2009) }.

\bibitem{PhysRevD.95.074009}
T.~Maji and D.~Chakrabarti, \emph{Transverse structure of a proton in a light-front quark-diquark model}, \href{https://doi.org/10.1103/PhysRevD.95.074009}{\emph{Phys. Rev. D} {\bfseries 95} (2017) 074009}.

\bibitem{chakrabarti2013generalized}
D.~Chakrabarti and C.~Mondal, \emph{Generalized parton distributions for the proton in ads/qcd}, {\emph{Physical Review D—Particles, Fields, Gravitation, and Cosmology} {\bfseries 88} (2013) 073006}.

\bibitem{Choudhary:2023unw}
P.~Choudhary, D.~Chakrabarti and C.~Mondal, \emph{{Flavor asymmetry of light sea quarks in proton: a light-front spectator model}}, \href{https://doi.org/10.1140/epjc/s10052-024-12985-2}{\emph{Eur. Phys. J. C} {\bfseries 84} (2024) 626} [\href{https://arxiv.org/abs/2312.01484}{{\ttfamily 2312.01484}}].

\bibitem{dove2023measurement}
J.~Dove, B.~Kerns, C.~Leung, R.~McClellan, S.~Miyasaka, D.~Morton et~al., \emph{Measurement of flavor asymmetry of the light-quark sea in the proton with drell-yan dimuon production in p+ p and p+ d collisions at 120 gev}, {\emph{Physical Review C} {\bfseries 108} (2023) 035202}.

\bibitem{towell2001improved}
R.~Towell, P.~McGaughey, T.~Awes, M.~Beddo, M.~Brooks, C.~Brown et~al., \emph{Improved measurement of the d{\={}}/{\=u} asymmetry in the nucleon sea}, {\emph{Physical Review D} {\bfseries 64} (2001) 052002}.

\bibitem{NA51:1994xrz}
{\scshape NA51} collaboration, \emph{{Study of the isospin symmetry breaking in the light quark sea of the nucleon from the Drell-Yan process}}, \href{https://doi.org/10.1016/0370-2693(94)90884-2}{\emph{Phys. Lett. B} {\bfseries 332} (1994) 244}.

\bibitem{meissner2009generalized}
S.~Mei{\ss}ner, A.~Metz and M.~Schlegel, \emph{Generalized parton correlation functions for a spin-1/2 hadron}, {\emph{Journal of High Energy Physics} {\bfseries 2009} (2009) 056}.

\bibitem{rajan2018lorentz}
A.~Rajan, M.~Engelhardt and S.~Liuti, \emph{Lorentz invariance and qcd equation of motion relations for generalized parton distributions and the dynamical origin of proton orbital angular momentum}, {\emph{Physical Review D} {\bfseries 98} (2018) 074022}.

\bibitem{Jain_2024}
S.~Jain, S.~Sharma and H.~Dahiya, \emph{Deciphering twist-3 chiral-even gpds in the light-front quark-diquark model}, \href{https://doi.org/10.1103/physrevd.110.094030}{\emph{Physical Review D} {\bfseries 110} (2024) }.

\bibitem{sharma2024exploring}
S.~Sharma and H.~Dahiya, \emph{Exploring twist-4 chiral-even gpds in the light-front quark-diquark model}, {\emph{Nuclear Physics B} {\bfseries 1001} (2024) 116522}.

\bibitem{brodsky1998quantum}
S.J.~Brodsky, H.-C.~Pauli and S.S.~Pinsky, \emph{Quantum chromodynamics and other field theories on the light cone}, {\emph{Physics Reports} {\bfseries 301} (1998) 299}.

\bibitem{harindranath1996introduction}
A.~Harindranath, \emph{An introduction to light-front dynamics for pedestrians}, {\emph{arXiv preprint hep-ph/9612244} (1996) }.

\bibitem{34lm-pr91}
X.~Luan and Z.~Lu, \emph{Twist-3 generalized parton distributions of sea quarks at zero skewness in the light-cone quark model}, \href{https://doi.org/10.1103/34lm-pr91}{\emph{Phys. Rev. D} (2025) }.

\bibitem{cite-key}
J.~Cruz-Martinez, T.~Hasenack, F.~Hekhorn, G.~Magni, E.R.~Nocera, T.R.~Rabemananjara et~al., \emph{Nnpdfpol2.0: a global determination of polarised pdfs and their uncertainties at next-to-next-to-leading order}, \href{https://doi.org/10.1007/JHEP07(2025)168}{\emph{Journal of High Energy Physics} {\bfseries 2025} (2025) 168}.

\bibitem{alekseev2010quark}
M.~Alekseev, V.Y.~Alexakhin, Y.~Alexandrov, G.~Alexeev, A.~Amoroso, A.~Austregesilo et~al., \emph{Quark helicity distributions from longitudinal spin asymmetries in muon--proton and muon--deuteron scattering}, {\emph{Physics Letters B} {\bfseries 693} (2010) 227}.

\bibitem{airapetian2004flavor}
A.~Airapetian, N.~Akopov, Z.~Akopov, M.~Amarian, V.~Ammosov, A.~Andrus et~al., \emph{Flavor decomposition of the sea-quark helicity distributions in the nucleon from semiinclusive deep inelastic scattering}, {\emph{Physical review letters} {\bfseries 92} (2004) 012005}.

\bibitem{jaffe1992chiral}
R.L.~Jaffe and X.~Ji, \emph{Chiral-odd parton distributions and drell-yan processes}, {\emph{Nuclear Physics B} {\bfseries 375} (1992) 527}.

\bibitem{guo2021novel}
Y.~Guo, X.~Ji and K.~Shiells, \emph{Novel twist-three transverse-spin sum rule for the proton and related generalized parton distributions}, {\emph{Nuclear Physics B} {\bfseries 969} (2021) 115440}.

\bibitem{Luan:2023lmt}
X.~Luan and Z.~Lu, \emph{{Generalized parton distributions of sea quark at zero skewness in the light-cone model}}, \href{https://doi.org/10.1140/epjc/s10052-023-11637-1}{\emph{Eur. Phys. J. C} {\bfseries 83} (2023) 504} [\href{https://arxiv.org/abs/2302.11278}{{\ttfamily 2302.11278}}].

\bibitem{diehl2001generalized}
M.~Diehl, \emph{Generalized parton distributions with helicity flip}, {\emph{The European Physical Journal C-Particles and Fields} {\bfseries 19} (2001) 485}.

\bibitem{BentzNJL}
W.~Bentz and T.~Ito, \emph{Applications of the nambu–jona‑lasinio model to hadron structure}, {\emph{Nucl. Phys. A (example)} (2008–2010) }.

\bibitem{boffi2003linking}
S.~Boffi, B.~Pasquini and M.~Traini, \emph{Linking generalized parton distributions to constituent quark models}, {\emph{Nuclear Physics B} {\bfseries 649} (2003) 243}.

\bibitem{brodsky2001light}
S.J.~Brodsky, D.S.~Hwang, B.-Q.~Ma and I.~Schmidt, \emph{Light-cone representation of the spin and orbital angular momentum of relativistic composite systems}, {\emph{Nuclear Physics B} {\bfseries 593} (2001) 311}.

\bibitem{bhattacharya2022burkhardt}
S.~Bhattacharya and A.~Metz, \emph{Burkhardt-cottingham-type sum rules for light-cone and quasi-pdfs}, {\emph{Physical Review D} {\bfseries 105} (2022) 054027}.

\end{thebibliography}\endgroup

\end{document}